\documentclass[prb,twocolumn,showpacs]{revtex4}
\usepackage{amsmath}
\usepackage{dcolumn}
\usepackage{graphicx}
\usepackage{bm}

\begin{document}

\title[Short title for running header]{Variational study of the ground state and spin dynamics of the spin-$\frac{1}{2}$ Kagome antiferromagnetic Heisenberg model and its implication on Hebertsmithite ZnCu$_{3}$(OH)$_{6}$Cl$_{2}$}
\author{Chun Zhang and Tao Li}
\affiliation{Department of Physics, Renmin University of China,
Beijing 100872, P.R.China}
\date{\today}

\begin{abstract}
We find that the best RVB state of the spin-$\frac{1}{2}$ Kagome antiferromagnetic Heisenberg model(spin-$\frac{1}{2}$ KAFH) is described by a $Z_{2}$ gapped mean field ansatz, which hosts a mean field spinon dispersion very different from that of the widely studied $U(1)$ Dirac spin liquid state. However, we find that the physical spin fluctuation spectrum calculated from the Gutzwiller projected RPA(GRPA) theory above such an RVB state is actually gapless and is almost identical to that above the $U(1)$ Dirac spin liquid state. We find that such a peculiar behavior can be attributed to the unique flat band physics on the Kagome lattice, which makes the mapping between the mean field ansatz and the RVB state non-injective. We find that the spin fluctuation spectrum of the spin-$\frac{1}{2}$ KAFH is not at all featureless, but is characterized by a prominent spectral peak at about $0.25J$ around the $\mathbf{M}$ point, which is immersed in a broad continuum extending to $2.7J$. Based on these results, we argue that the spectral peak below 2 meV in the inelastic neutron scattering(INS) spectrum of Hebertsmithite ZnCu$_{3}$(OH)$_{6}$Cl$_{2}$, which has been attributed to the contribution of Cu$^{2+}$ impurity spins occupying the Zn$^{2+}$ site, should rather be understood as the intrinsic contribution from the Kagome layer. We propose to verify such a picture by measuring the Knight shift on the Cu site, rather than the O site, which is almost blind to the spin fluctuation at the $\mathbf{M}$ point as a result of the strong antiferromagnetic correlation between nearest neighboring spins on the Kagome lattice. 
\end{abstract}


\maketitle
\section{Introduction}
The spin-$\frac{1}{2}$ Kagome antiferromagnetic Heisenberg model with nearest-neighboring exchange (spin-$\frac{1}{2}$ KAFH) is extensively studied in the last three decades in quest of quantum spin liquid\cite{Elser,Chalker,Leung,Young,Lecheminant,Sindzingre,Nakano,Lauchli,Series,Vidal,Singlet,Mila,Auerbach,Poilblanc,Sheng,He1,Changlani}. While it is generally believed that the ground state of the spin-$\frac{1}{2}$ KAFH is a quantum spin liquid, it is strongly debated what is the exact nature of such a novel quantum state of matter. Variational studies based on the resonating valence bond(RVB) theory have accumulated extensive evidence for a gapless $U(1)$ Dirac spin liquid scenario\cite{Hastings,Ran,Iqbal1,Iqbal2,Iqbal3}, which is also implied by some recent studies using other numerical approaches\cite{Liao,He,Jiang,Zhu}. On the other hand, a gapped $Z_{2}$ spin liquid ground state has been claimed by many other studies\cite{Jiang1,Yan,Depenbrock,Jiang2,Kolley,Gong,Wen}. Clearly, a study of the spin fluctuation spectrum of the system is the most straightforward way to resolve the controversy among these ground-state-oriented studies.

On the experimental side, Hebertsmithite ZnCu$_{3}$(OH)$_{6}$Cl$_{2}$ is argued to be an ideal realization of the spin-$\frac{1}{2}$ KAFH, with possible Cu$^{2+}$ impurity spins occupying the Zn$^{2+}$ sites between the Kagome layer\cite{Mendels,Helton}. Inelastic neutron scattering(INS) measurement on Hebertsmithite ZnCu$_{3}$(OH)$_{6}$Cl$_{2}$ finds that the spin fluctuation spectrum of the system is characterized by a broad and featureless continuum above 2 meV.  Below 2 meV, a broad peak emerges around the M point of the Brillouin zone\cite{Han}. This peak is believed to be contributed by the Cu$^{2+}$ impurity spins between the Kagome layers. Such a scenario is supported by a later NMR measurement on the system\cite{Fu}, which implies a finite gap of $\Delta= 0.03J \sim 0.07J$ in the intrinsic spin fluctuation spectrum of the Kagome layer. A spin gap of similar size is also reported in the NMR study of a related Kagome material ZnCu$_{3}$(OH)$_{6}$FBr\cite{Shi}. However, it is puzzling why the low energy spectral peak can exhibit such a strong momentum dependence as observed in INS measurement, if it is indeed contributed by impurity spins. More recently, a refined NMR measurement on Hebertsmithite ZnCu$_{3}$(OH)$_{6}$Cl$_{2}$ finds that the intrinsic spin fluctuation spectrum of the Kagome layer is actually gapless\cite{Mendels}. 
 
Within the RVB theory framework, attempts have been made to reconcile the discrepancy between the different theories regarding the spin excitation gap of the spin-$\frac{1}{2}$ KAFH. Using projective symmetry group(PSG) analysis, people find that a gapped $Z_{2}$ spin liquid state can indeed be realized in the vicinity of the $U(1)$ Dirac spin liquid state if one introduce longer-ranged RVB parameters\cite{Lu,Wen1}. However, VMC calculations along this line generate controversial results\cite{Iqbal1,Iqbal2,Iqbal3,Tao1,Iqbal4,Iqbal5}. In a recent work\cite{Tao2}, we find that the mapping between the mean field ansatz and the RVB state becomes non-injective around the $U(1)$ Dirac spin liquid state as a result of an unique flat band physics on the Kagome lattice. More specifically, we find that the $U(1)$ Dirac spin liquid state with only nearest-neighboring RVB parameter\cite{Hastings,Ran} can be generated from a continuous family of gauge inequivalent RVB mean field ansatz. The mean field spinon dispersion within this family changes violently as we tune the RVB parameter on the second and the third neighboring bonds. The implication of such a peculiar behavior is twofold. First, it implies that the optimization of the RVB parameters around the $U(1)$ Dirac spin liquid state is a rather subtle practice. In particular, we should include simultaneously the second and the third neighbor RVB parameters in the variational optimization. Second, it implies that the spin fluctuation spectrum calculated at the mean field level is unphysical. We must go beyond the mean field theory to resolve the ambiguity in the spin fluctuation spectrum. 

With these considerations in mind, we have performed a systematic variational Monte Carlo(VMC) study on the ground state and spin fluctuation spectrum of the spin-$\frac{1}{2}$ KAFH within the RVB theory framework. More specifically, we have performed a large scale variational optimization of RVB state for the spin-$\frac{1}{2}$ KAFH with both the second and the third neighboring RVB parameters. We then calculated the spin fluctuation spectrum of the system with the Gutzwiller projected RPA(GRPA) theory, which has been proved to be rather successful in the study of dynamical property of strongly correlated electron systems\cite{Li,Li1,Piazza,Mei1,Ferrari,Ferrari1,Becca,Ido,Becca1}. 

We find that the best RVB state of the spin-$\frac{1}{2}$ KAFH is described by a $Z_{2}$ gapped mean field ansatz, which hosts a mean field spinon dispersion very different from that of the $U(1)$ Dirac spin liquid state. However, when we go beyond the mean field description, we find that spin fluctuation spectrum above the optimized RVB state is actually gapless and is almost identical to that above the $U(1)$ Dirac spin liquid state. Unlike the mean field prediction, we find that the spin fluctuation spectrum of the spin-$\frac{1}{2}$ KAFH is not at all featureless, but is characterized by a prominent spectral peak at about $0.25J$ around the M point of the Brillouin zone. We find that such an in tense spectral peak is immersed in a gapless spin fluctuation continuum, which extends to an energy as high as $2.7J$. We find that the spectral characteristic predicted by the GRPA theory agrees well with the prediction of recent dynamical DMRG simulation on the spin-$\frac{1}{2}$ KAFH\cite{Zhu}. 

We have compared our variational spin fluctuation spectrum with the INS result on Hebertsmithite ZnCu$_{3}$(OH)$_{6}$Cl$_{2}$. Since the intense spectral peak around the M point is such a robust feature in the spin fluctuation spectrum of the spin-$\frac{1}{2}$ KAFH, we argue that the spectral peak below 2 meV in the INS spectrum of Hebertsmithite ZnCu$_{3}$(OH)$_{6}$Cl$_{2}$ should be attributed to the intrinsic spin fluctuation of the Kagome layer, rather than the contribution from the Cu$^{2+}$ impurity spins between the Kagome layers. This assignment resolve immediately the puzzle concerning the strong momentum dependence of this low energy spectral peak. It also implies that the ground state of Hebertsmithite ZnCu$_{3}$(OH)$_{6}$Cl$_{2}$ is much closer to magnetic ordering instability toward the $\mathbf{q}=0$ order than we thought before. We propose to verify such a picture by measuring Knight shift on the Cu site, rather than the O site, which is blind to the spin fluctuation at the M point as a result of the strong antiferromagnetic correlation between nearest neighboring spins on the Kagome lattice.
 
The paper is organized as follows. The theoretical formalism of this work is presented in the Sec.II. In this section, we will introduce the RVB theory for the spin-$\frac{1}{2}$ KAFH in both its $U(1)$ and $Z_{2}$ form and discuss the subtleties of applying the RVB theory on the Kaogme lattice caused by its unique flat band physics. We then present a Gutzwiller projected RPA(GRPA) theory for the spin fluctuation spectrum on the RVB ground state. The numerical result generated from the above RVB theory is presented in Sec.III. In this section, we will show that the optimized RVB state for the spin-$\frac{1}{2}$ KAFH is described by a gapped $Z_{2}$ mean field ansatz. We then show that contrary to the prediction of the mean field theory, the spin fluctuation spectrum above the optimized RVB state is actually gapless and is almost identical to that above the $U(1)$ Dirac spin liquid state. We then compare the theoretical prediction and the INS result on Hebertsmithite ZnCu$_{3}$(OH)$_{6}$Cl$_{2}$. In particular, we will show that the low energy peak around the M point observed in the INS spectra of Hebertsmithite ZnCu$_{3}$(OH)$_{6}$Cl$_{2}$ should be attributed to intrinsic spin fluctuation of the Kagome layers. The last section of the paper is devoted to a conclusion of the results and a discussion of some remianing problems.

\section{An RVB theory of the spin-$\frac{1}{2}$ KAFH}
\subsection{The RVB ground state of spin-$\frac{1}{2}$ KAFH}
The spin-$\frac{1}{2}$ KAFH studied in this work has the Hamiltonian
\begin{equation}
H_{J}=J\sum_{<i,j>}\mathbf{S}_{i}\cdot\mathbf{S}_{j}.
\end{equation}
The sum is over nearest-neighboring bonds of the Kagome lattice. To describe the spin liquid ground state of the system in the RVB scheme, we introduce Fermionic slave particle $f_{\alpha}$ and represent the spin operator as $\mathbf{S}=\frac{1}{2}\sum_{\alpha,\beta}f^{\dagger}_{\alpha}\bm{\sigma}_{\alpha,\beta}f_{\beta}$. Such a representation is exact when the slave Fermion satisfy the constraint $\sum_{\alpha}f^{\dagger}_{\alpha}f_{\alpha}=1$. The RVB state is generated from Gutzwiller projection of  BCS-type mean field ground state
\begin{equation}
|\mathrm{RVB}\rangle=\mathrm{P_{G}}|\mathrm{BCS}\rangle.
\end{equation}
Here $\mathrm{P_{G}}$ denotes the Gutzwiller projection into the singly occupied subspace. $|\mathrm{BCS}\rangle$ is the ground state of the following BCS-type Hamiltonian
\begin{equation}
H_{MF}=\sum_{i,j}\psi_{i}^{\dagger}U_{i,j}\psi_{j}.
\end{equation}
Here $\psi_{i}=\left(\begin{array}{c}f_{i,\uparrow}\\f^{\dagger}_{i,\downarrow}\end{array}\right)$, 
$U_{i,j}=\left(\begin{array}{cc}\chi_{i,j} & \Delta^{*}_{i,j} \\\Delta_{i,j} & -\chi^{*}_{i,j}\\ \end{array}\right)$. $\chi_{i,j}$ and $\Delta_{i,j}$ denote the RVB parameters in the hopping and pairing channel. The structure information of the RVB state is encoded in the mean field Hamiltonian $H_{MF}$, which is usually called a mean field ansatz of the RVB state.

We note that the RVB state so constructed is invariant when we perform a $SU(2)$ gauge transformation of the form $U_{i,j}\rightarrow G^{\dagger}_{i}U_{i,j}G_{j}$ on the RVB parameter $U_{i,j}$, in which $G_{i}$ is a site-dependent $SU(2)$ matrix\cite{Wen1}. Thus, to generate a symmetric spin liquid state, the RVB order parameter $U_{i,j}$ should be invariant under the symmetry operations only up to a $SU(2)$ gauge transformation. The structure of the RVB state can thus be classified by the gauge inequivalent way to choose such a gauge transformation\cite{Wen1}. For example, in a $Z_{2}$ spin liquid state, the translational symmetry can be realized either by assuming a translational invariant RVB mean filed ansatz, or an RVB mean field ansatz that differs by a $Z_{2}$ gauge transformation from the translated ansatz. Here we only consider $Z_{2}$ spin liquid state of the second type, which can have a smooth connection with the $U(1)$ Dirac spin liquid state studied in Ref.[\onlinecite{Ran}]. At the same time, we will keep RVB order parameters $U_{i,j}$ up to the third neighboring bonds.

The mean field ansatz of the $U(1)$ and the $Z_{2}$ spin liquid state studied in this work is illustrated in Fig. 1. The yellow parallelogram denotes the unit cell of the Kagome lattice, with $\mathrm{\mathbf{a}}_{1}$ and $\mathrm{\mathbf{a}}_{2}$ as its two basis vectors. The blue, yellow and pink lines denote the first, second and the third neighboring bonds on the Kagome lattice. For the spin liquid state studied here, $U_{i,j}$ is translational invariant along the $\mathrm{\mathbf{a}}_{2}$ direction, but will change sign when translated in the $\mathrm{\mathbf{a}}_{1}$ direction by one lattice constant, if the cell index of site $i$ and $j$ in the $\mathrm{\mathbf{a}}_{2}$ direction differ by an odd number.

\begin{figure}
\includegraphics[width=7cm,height=5.5cm]{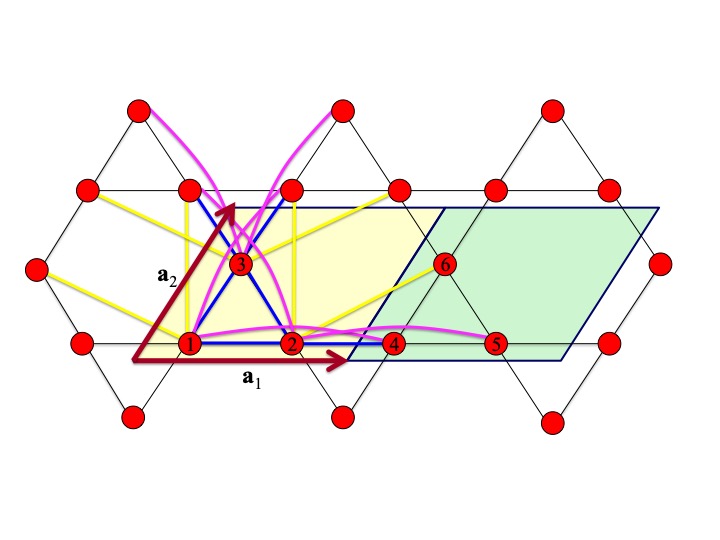}
\includegraphics[width=5.5cm,height=4.5cm]{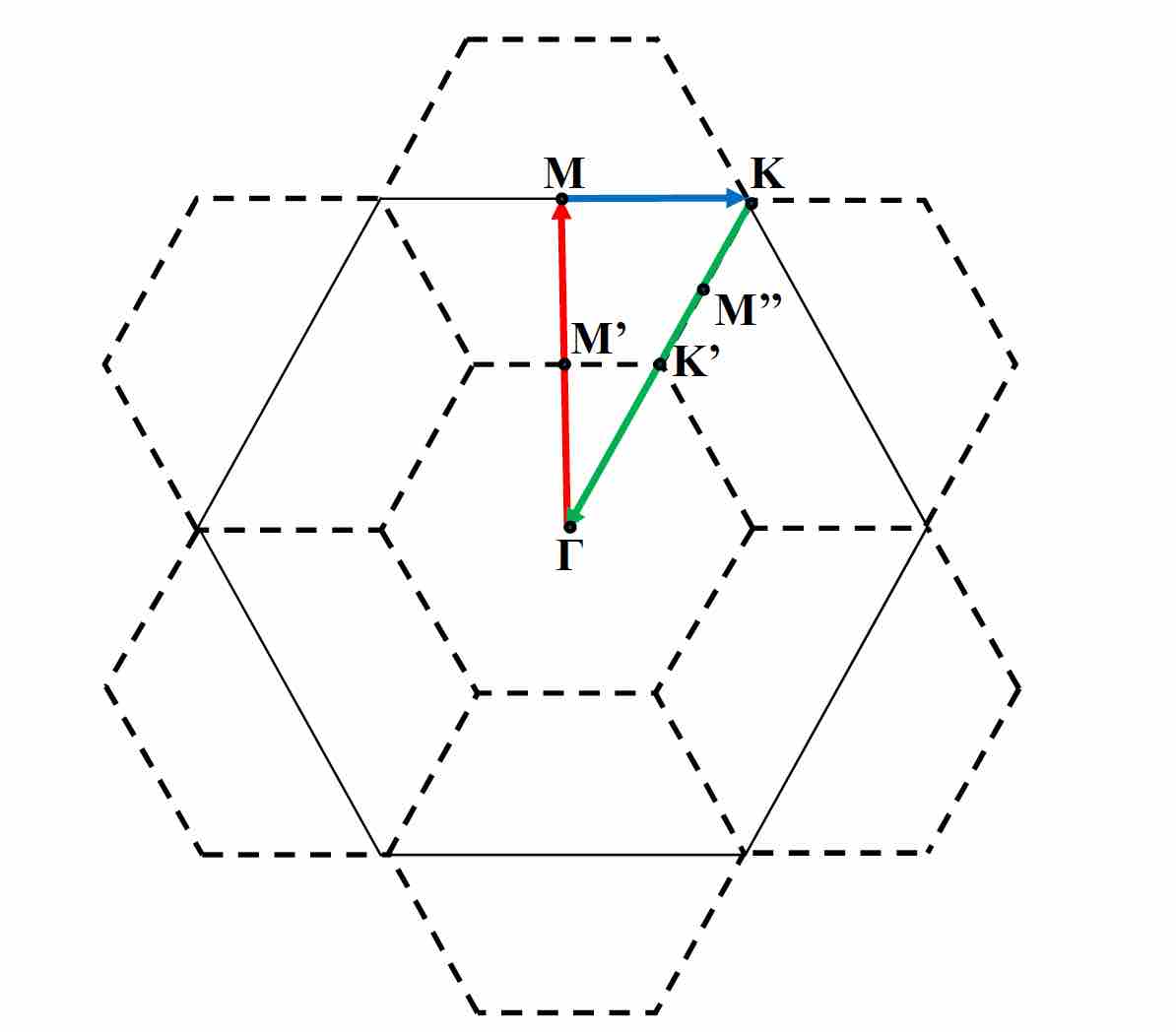}
\caption{(Upper panel) Illustration of the mean field ansatz of the $U(1)$ and the $Z_{2}$ spin liquid state studied in this paper. The yellow 
parallelogram denotes the unit cell of the Kagome lattice, with $\mathrm{\mathbf{a}}_{1}$ and $\mathrm{\mathbf{a}}_{2}$ as its two basis vectors.
 The spinon unit cell is doubled in the $\mathrm{\mathbf{a}}_{1}$ direction and contains six sites(site $\mu=1,....,6$ shown in the figure). The blue, yellow and pink lines denote the first, second and the third neighboring bonds of the Kagome lattice. $U_{i,j}$ is translational invariant along the $\mathrm{\mathbf{a}}_{2}$ direction, but will change sign when translated in the $\mathrm{\mathbf{a}}_{1}$ direction by one lattice constant, if the cell indices in the $\mathrm{\mathbf{a}}_{2}$ direction of site $i$ and $j$ differ by an odd number. (Lower panel)Illustration of the path in the momentum space($\Gamma-\mathbf{M}'-\mathbf{M}-\mathbf{K}-\mathbf{M}''-\Gamma$) along which the spin fluctuation spectrum is calculated. The elementary hexagons plotted in dashed line denote the Brillouin zones of the Kagome lattice.} \label{fig1}
\end{figure}

For the $U(1)$ spin liquid state, the RVB parameters take the form
\begin{eqnarray}
U_{i,j}=\left\{\begin{aligned}
                -s_{i,j}\ \tau_{3}& & \mathrm{first\ neighbor}\\  
                -s_{i,j}\ \rho\tau_{3}& & \mathrm{second\ neighbor}\\
                -s_{i,j}\ \eta\tau_{3}& & \mathrm{third\ neighbor}
\end{aligned}  
\right.
\end{eqnarray}
Here a chemical potential term is implicitly assumed to enforce the half-filling condition on the Fermion number. $\tau_{3}=\left(\begin{array}{cc}1 & 0 \\0 & -1\end{array}\right)$ is a Pauli matrix. $\rho$ and $\eta$ are two real variational parameters. $s_{i,j}=\pm1$ is introduced to generate the sign change when we translate $U_{i,j}$ in the $\mathrm{\mathbf{a}}_{1}$ direction. They equal to 1 on the blue, yellow and pink bonds shown in Fig. 1. When $\rho=\eta=0$, $U_{i,j}$ reduces to the mean field ansatz of the $U(1)$ Dirac spin liquid state first studied in Ref.[\onlinecite{Ran}]. We will refer to such a state as $U(1)$-NN state for brevity below. 

In a previous work\cite{Tao2}, we have shown that the mapping between the mean field ansatz and the RVB state becomes non-injective when $\rho=\eta$. More specifically, for $-0.6\leq\rho=\eta\leq0.27$, the mean field ground state corresponding to the ansazt Eq.(4) is independent of the value of $\rho$.  Such a peculiar behavior can be understood if we rewrite $H_{MF}$ as
\begin{equation}
H_{MF}=H_{1}+H_{\rho}+H_{\eta},
\end{equation} 
in which $H_{1}, H_{\rho}$ and $H_{\eta}$ denote the part of $H_{MF}$ contributed by spinon hopping between the first, second and the third neighboring sites. It is then easy to check that  
\begin{equation}
[H_{1},H_{\rho}+H_{\eta}]=0,
\end{equation}
when $\rho=\eta$. The eigenstate of the mean field Hamiltonian thus does not depend on the value of $\rho$ when $\rho=\eta$.

\begin{figure}
\includegraphics[width=7cm,height=5.5cm]{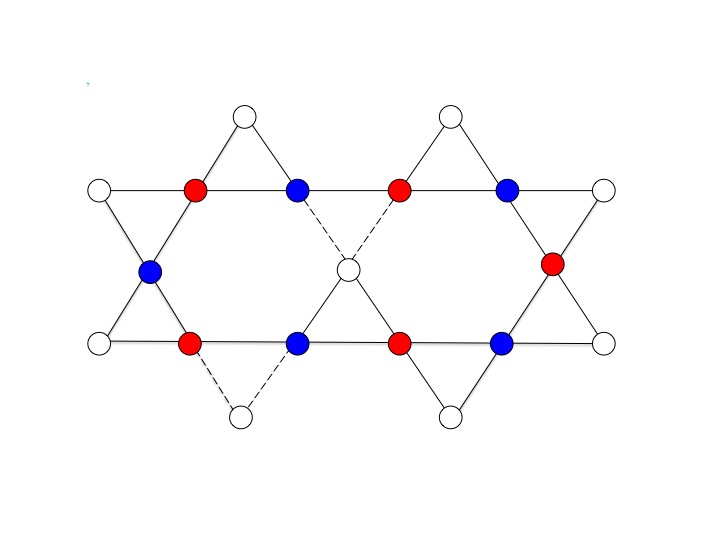}
\caption{Illustration of the localized Wannier orbital of the flat band of the mean field ansatz Eq.(4). The wave function amplitude on the red, blue and white sites are +1, -1 and 0. The hopping integral on the dashed bonds has an additional minus sign as a result of the factor $s_{i,j}$ in Eq.(4). The hopping amplitudes from the red and the blue sites to any white site add to zero when $\rho=\eta$.} \label{fig2}
\end{figure}

The peculiarity of $H_{MF}$ discussed above is deeply related to the unique flat band physics on the Kagome lattice. It is well known that there is a flat band (with an eigenvalue of 2) in the mean field spinon dispersion\cite{Ran} of the $U(1)$-NN state(i.e., when $\rho=\eta=0$). The origin of such a flat band can be traced back to the destructive interference between the hopping amplitudes out of a localized Wannier orbital of the form shown in Fig.2. Interestingly, one find that such a destructive interference remains effective in the presence of the second and the third neighboring hopping terms, provided that $\rho=\eta$.  A nonzero $\rho=\eta$ thus only shift the eigenvalue of the flat band, but does not change its wave function. In the whole range of $\rho(=\eta)\in [-0.6,0.27]$, one find that the occupied state is independent of the value of $\rho=\eta$. Thus the $U(1)$-NN state originally studied in Ref.[\onlinecite{Ran}] can actually be generated from a continuously family of gauge inequivalent RVB mean field ansatz. Such a peculiarity not only complicates the optimization of the RVB parameters around the $U(1)$-NN state, but also implies that the spin fluctuation spectrum calculated at the mean field level is unphysical, in we insist on relating the excitation characteristic of a quantum system to its ground state structure.

For the $Z_{2}$ spin liquid state, the RVB parameters take the form
\begin{eqnarray}
U_{i,j}=\left\{\begin{aligned}
                -\mu \vec{n}_{\phi_{1}} \cdot \vec{\tau} & & \mathrm{on-site} \\
                -s_{i,j}\ \tau_{3}& & \mathrm{first\ neighbor}\\  
                -s_{i,j}\ \rho  \vec{n}_{\phi_{2}} \cdot \vec{\tau}& & \mathrm{second\ neighbor}\\
                -s_{i,j}\ \eta  \vec{n}_{\phi_{3}} \cdot \vec{\tau}& & \mathrm{third\ neighbor}
\end{aligned}  
\right.
\end{eqnarray}
Here $\vec{\tau}=(\tau_{1},\tau_{2},\tau_{3})$ are the Pauli matrices, $\vec{n}_{\phi}=(\sin\phi,0,\cos\phi)$ is a unit vector in the $\tau_{1}-\tau_{3}$ plane.
 $\mu,\rho,\eta$ and $\phi_{1,2,3}$ are six real variational parameters of the $Z_{2}$ spin liquid state. It can be easily checked that spin liquid state generated from the $U(1)$ and the $Z_{2}$ ansatz respect all physical symmetry of the spin-$\frac{1}{2}$ KAFH. At the same time, the $Z_{2}$ spin liquid state reduces to the $U(1)$ spin liquid state when $\phi_{1,2,3}=N\pi$, in which $N$ is an arbitrary integer. The mapping between the RVB parameters and the spin liquid state becomes non-injective when $\rho=\eta$ and $\phi_{1}=\phi_{2}=\phi_{3}=\pi$.

\subsection{The spin fluctuation spectrum of the spin-$\frac{1}{2}$ KAFH above the RVB ground state}
 The spin fluctuation spectrum of the system can be extracted from the dynamical spin susceptibility defined below 
\begin{equation}
\bm{\chi}^{i,j}(\mathbf{q},\tau)=-\langle \ T_{\tau} \mathrm{S}^{i}(\mathbf{q},\tau)\  \mathrm{S}^{j}(-\mathbf{q},0) \rangle.
 \end{equation}
Here 
\begin{equation}
\mathrm{S}^{i}(\mathbf{q})=\frac{1}{2}\sum_{\mathbf{k},\alpha,\beta,\mu}\  e^{i\mathbf{q}\cdot\bm{\delta}_{\mu}} \ f^{\dagger}_{\mathbf{k+q},\mu,\alpha}\sigma^{i}_{\alpha,\beta}f_{\mathbf{k},\mu,\beta},
\end{equation}
is the $i$-th component of spin density operator at momentum $\mathbf{q}$. $\mu=1,....,6$ denotes the index of the six sublattices in the spinon unit cell. $\bm{\delta}_{\mu}$ is the displacement of the $\mu$-th sublattice with respect to the origin of the spinon unit cell. Since the ground state of the system is spin rotational symmetric, we will concentrate on the fluctuation of the z-component of the spin density operator, which can be expressed in terms of the Nambu spinor as follows
\begin{equation}
\mathrm{S}^{z}(\mathbf{q})=\frac{1}{2}\sum_{\mathbf{k},\mu} \ e^{i\mathbf{q}\cdot\bm{\delta}_{\mu}} \  \psi^{\dagger}_{\mathbf{k+q},\mu}\psi_{\mathbf{k},\mu},
\end{equation}
in which $\psi_{\mathbf{k},\mu}=\left(\begin{array}{c}f_{\mathbf{k},\mu,\uparrow}\\f^{\dagger}_{-\mathbf{k},\mu,\downarrow}\end{array}\right)$ is the Nambu spinor. At the mean field level, the dynamical spin susceptibility is given by
\begin{equation}
\bm{\chi}_{0}^{z,z}(\mathbf{q},\tau)=\frac{1}{4}\sum_{\mathbf{k},\mu,\nu} e^{i\mathbf{q}\cdot(\bm{\delta}_{\mu}-\bm{\delta}_{\nu})}  \mathbf{Tr}[ \mathbf{G}_{\nu\mu}(\mathbf{k+q},-\tau)\mathbf{G}_{\mu,\nu}(\mathbf{k},\tau)],
\end{equation}
in which $\mathbf{G}_{\mu,\nu}(\mathbf{k},\tau)=-\langle \ T_{\tau} \psi_{\mathbf{k},\mu}(\tau) \psi^{\dagger}_{\mathbf{k},\nu}(0)\rangle$ is the spinon Green's function calculated at the mean field level.

Anticipating the inadequacy of the mean field theory, in particular the subtleties of the RVB mean field theory on the Kagome lattice related to its unique flat band physics, we present in the following a Gutzwiller projected RPA(GRPA) theory of the spin fluctuation spectrum for the spin-$\frac{1}{2}$ KAFH. The essence of such a theory is to diagonalize the  Heisenberg Hamiltonian in a subspace spanned by the Gutzwiller projected mean field excited state\cite{Li}. These Gutzwiller projected mean field excited states are connected to the RVB ground state through the operation of the spin density operator $\mathrm{S}^{z}(\mathbf{q})$. Since $[\mathrm{S}^{z}(\mathbf{q}),\mathrm{P_{G}}]=0$, we have
\begin{eqnarray}
\mathrm{S}^{z}(\mathbf{q})|\mathrm{RVB}\rangle&=&\mathrm{S}^{z}(\mathbf{q})\mathrm{P_{G}}|\mathrm{BCS}\rangle=\mathrm{P_{G}}\mathrm{S}^{z}(\mathbf{q})|\mathrm{BCS}\rangle\nonumber\\
&=&\sum_{\mathbf{k},m,n}\mathrm{P_{G}}\phi_{\mathbf{k},m,n}(\mathbf{q})\gamma^{\dagger}_{\mathbf{k+q},m,\uparrow}\gamma^{\dagger}_{-\mathbf{k},n,\downarrow}|\mathrm{BCS}\rangle\nonumber\\
&=&\sum_{\mathbf{k},m,n}\phi_{\mathbf{k},m,n}(\mathbf{q})\mathrm{P_{G}}\gamma^{\dagger}_{\mathbf{k+q},m,\uparrow}\gamma^{\dagger}_{-\mathbf{k},n,\downarrow}|\mathrm{BCS}\rangle\nonumber\\
&=&\sum_{\mathbf{k},m,n}\phi_{\mathbf{k},m,n}(\mathbf{q})|\mathbf{k,q},m,n\rangle,
\end{eqnarray}
in which 
\begin{equation}
\phi_{\mathbf{k},m,n}(\mathbf{q})=\frac{1}{2}\sum_{\mu}e^{i\mathbf{q}\cdot\bm{\delta}_{\mu}}(v^{*}_{\mathbf{k+q},\mu,m}u_{\mathbf{k},\mu,n}-u^{*}_{\mathbf{k+q},\mu,m}v_{\mathbf{k},\mu,n}).
\end{equation}
 $|\mathbf{k,q},m,n\rangle=\mathrm{P_{G}}\gamma^{\dagger}_{\mathbf{k+q},m,\uparrow}\gamma^{\dagger}_{-\mathbf{k},n,\downarrow}|\mathrm{BCS}\rangle$ is a Gutzwiller projected mean field excited state generate by $\mathrm{S}^{z}(\mathbf{q})$. $m$ and $n$ are the indices of the quasiparticle eigenstate. The quasiparticle operator $\gamma_{\mathbf{k},m,\alpha}$ is related to the bare spinon operator $f_{\mathbf{k},\mu,\alpha}$ through the Bogoliubov transformation
\begin{equation}
\left(\begin{array}{c}f_{\mathbf{k},\mu,\uparrow} \\f^{\dagger}_{-\mathbf{k},\mu,\downarrow} \end{array}\right)=\left(\begin{array}{cc}\mathbf{u_{k}} & -\mathbf{v_{k}} \\ \mathbf{v_{k}} & \mathbf{u_{k}}\end{array}\right)\left(\begin{array}{c}\gamma_{\mathbf{k},m,\uparrow} \\\gamma^{\dagger}_{-\mathbf{k},m,\downarrow} \end{array}\right),
\end{equation}
in which $\mathbf{u_{k}}$ and $\mathbf{v_{k}}$ are $6\times6$ matrix with $u_{\mathbf{k},\mu,m}$ and $v_{\mathbf{k},\mu,m}$ as their matrix element. It can be shown that if both $\chi_{i,j}$ and $\Delta_{i,j}$ are real and symmetric, then $\mathbf{u_{-k}}=\mathbf{u^{*}_{k}}$, $\mathbf{v_{-k}}=\mathbf{v^{*}_{k}}$.

We thus choose the subspace spanned by $|\mathbf{k,q},m,n\rangle=\mathrm{P_{G}}\gamma^{\dagger}_{\mathbf{k+q},m,\uparrow}\gamma^{\dagger}_{-\mathbf{k},n,\downarrow}|\mathrm{BCS}\rangle$ as our working subspace and diagonalize $H_{J}$ in this subspace to construct a variational approximation of the spin fluctuation spectrum for the spin-$\frac{1}{2}$ KAFH. The Gutzwiller projected mean field excited state $|\mathbf{k,q},m,n\rangle$ is in general not orthonormal, we thus have to solve a generalized eigenvalue problem of the form
\begin{equation}
\mathbf{H}\bm{\varphi}_{i}=\lambda_{i} \mathbf{O}\bm{\varphi}_{i}.
\end{equation} 
Here the element of the Hamiltonian matrix $\mathbf{H}$ and the overlap matrix $\mathbf{O}$ are given by
\begin{eqnarray}
\mathbf{H}_{\mathbf{k}',m',n'; \mathbf{k},m,n}&=&\langle \mathbf{k}',\mathbf{q},m',n'|H_{J}| \mathbf{k},\mathbf{q},m,n\rangle\nonumber\\
\mathbf{O}_{\mathbf{k}',m',n'; \mathbf{k},m,n}&=&\langle \mathbf{k}',\mathbf{q},m',n'| \mathbf{k},\mathbf{q},m,n\rangle,
\end{eqnarray}
These matrix elements can be evaluated with a highly efficient re-weighting technique in VMC simulation\cite{Li}.
$\bm{\varphi}_{i}$ denotes the generalized eigenvector of Eq.(15) with eigenvalue $\lambda_{i}$. It is normalized as follows
\begin{equation}
\bm{\varphi}^{\dagger}_{i}\mathbf{O}\bm{\varphi}_{j}=\delta_{i,j}.
\end{equation}
The variational spin fluctuation spectrum calculated in the above subspace is given by
\begin{equation}
S(\mathbf{q},\omega)=\frac{1}{N}\sum_{i}|\bm{\phi}^{\dagger}\mathbf{O}\bm{\varphi}_{i}|^{2} \delta(\omega-(\lambda_{i}-E_{g})),
\end{equation}
in which $\bm{\phi}$ is a vector with $\phi_{\mathbf{k},m,n}$ as its components, $E_{g}$ is the variational ground state energy of the RVB state, $N$ denotes the number of lattice site in the system. 

In recent years, such a variational theory has been applied successfully in the study of dynamical properties of several strongly correlated electron systems\cite{Li,Li1,Piazza,Mei1,Ferrari,Ferrari1,Becca,Ido,Becca1}. The internal consistency of the theory can be seen from the fact that the spin fluctuation spectrum so constructed satisfies the momentum-resolved sum rule of the form
\begin{equation}
\int_{0}^{\infty} d\omega S(\mathbf{q},\omega)=S(\mathbf{q})=\frac{1}{N}\langle \mathrm{S}^{z}(\mathbf{q}) \mathrm{S}^{z}(\mathbf{-q})\rangle.
\end{equation}
Here $S(\mathbf{q})$ denotes the static spin structure factor calculated on the RVB ground state. 
\begin{figure}
\includegraphics[width=6cm,angle=0]{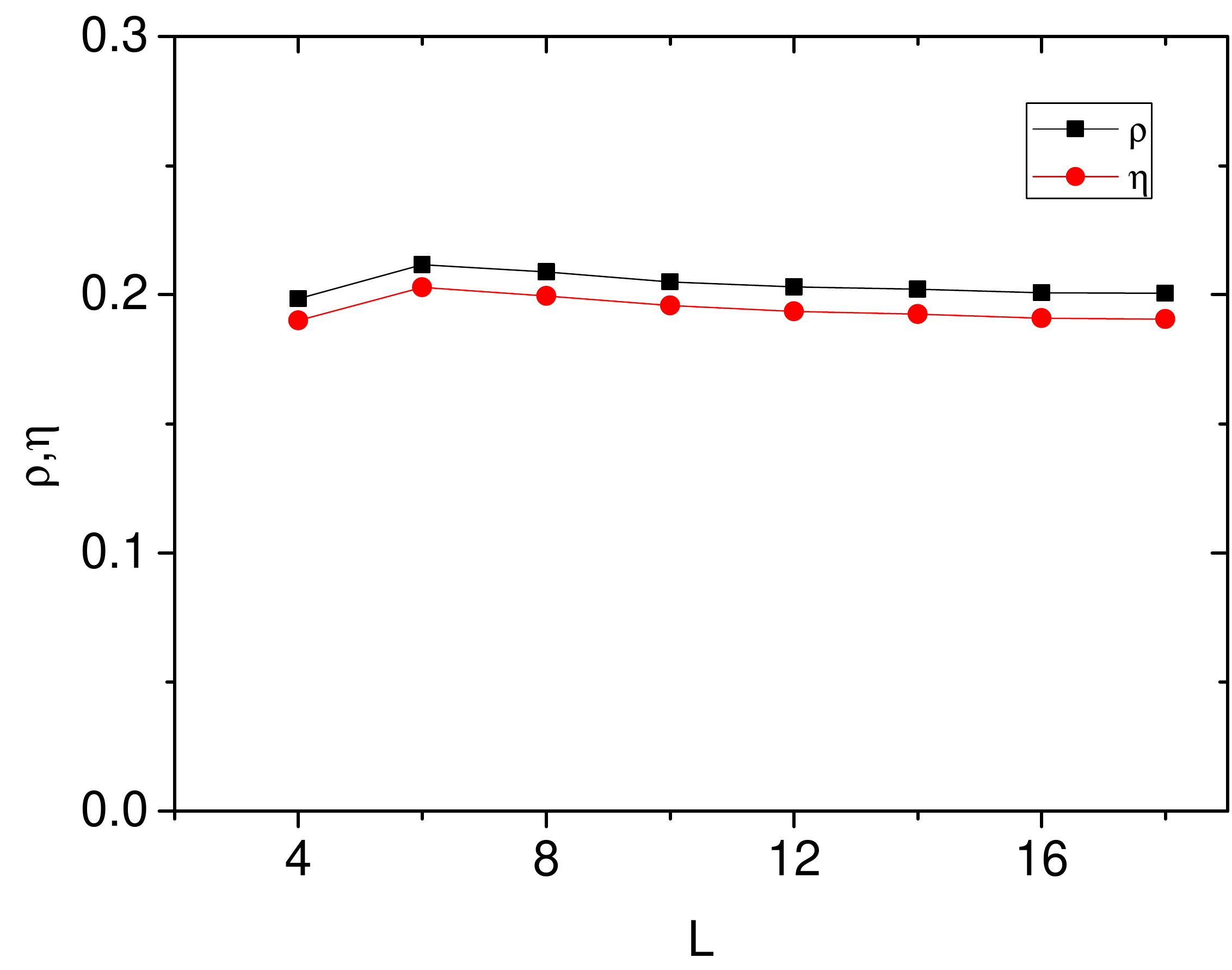}
\caption{The optimized value of the RVB parameters in the $U(1)$ spin liquid state. The optimized RVB parameters are found to be very close the non-injective line $\rho=\eta$, where the mapping between the mean field ansatz and the RVB state becomes singular.} \label{fig2}
\end{figure}

\begin{figure}
\includegraphics[width=6cm,angle=0]{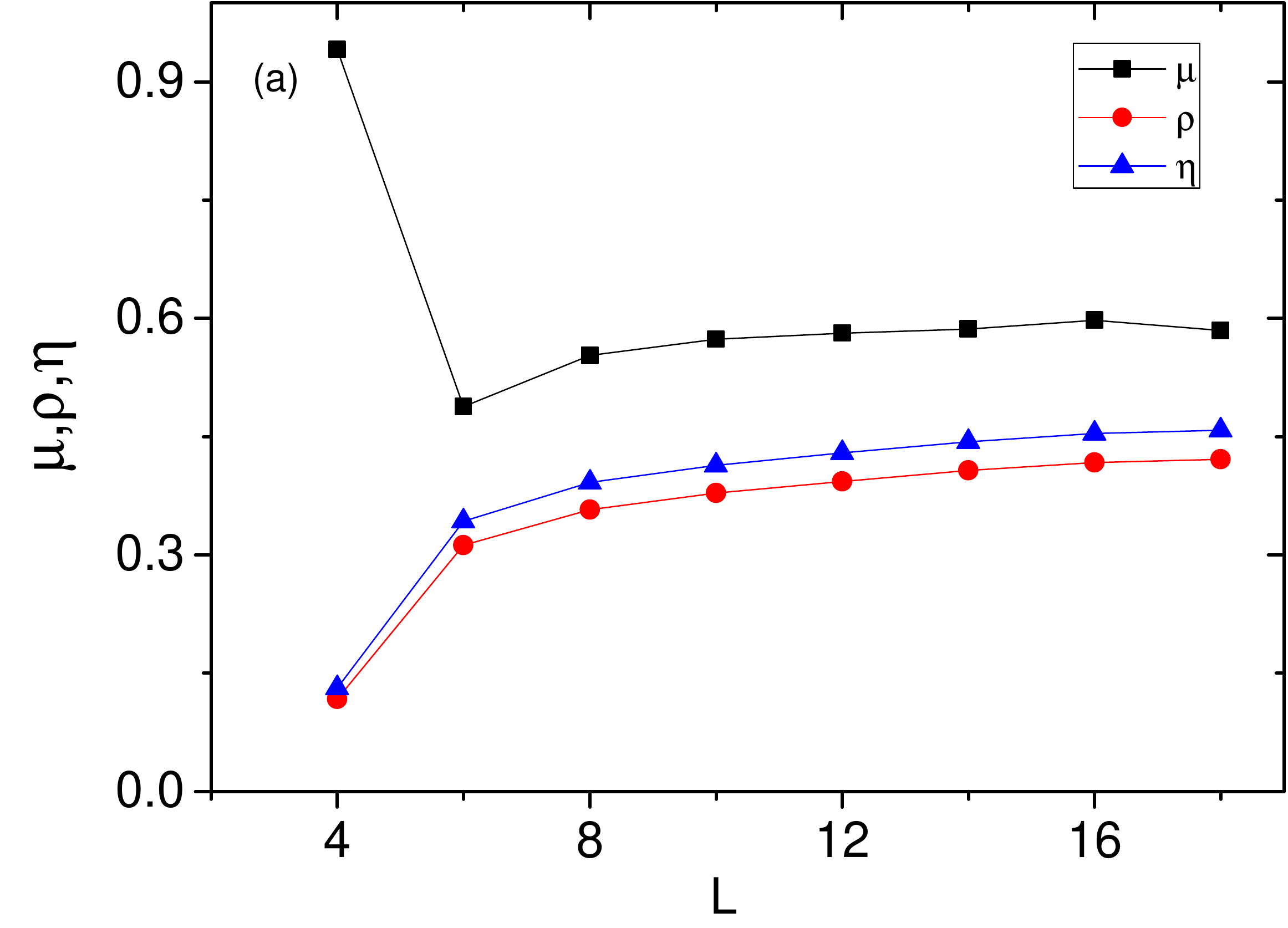}
\includegraphics[width=6cm,angle=0]{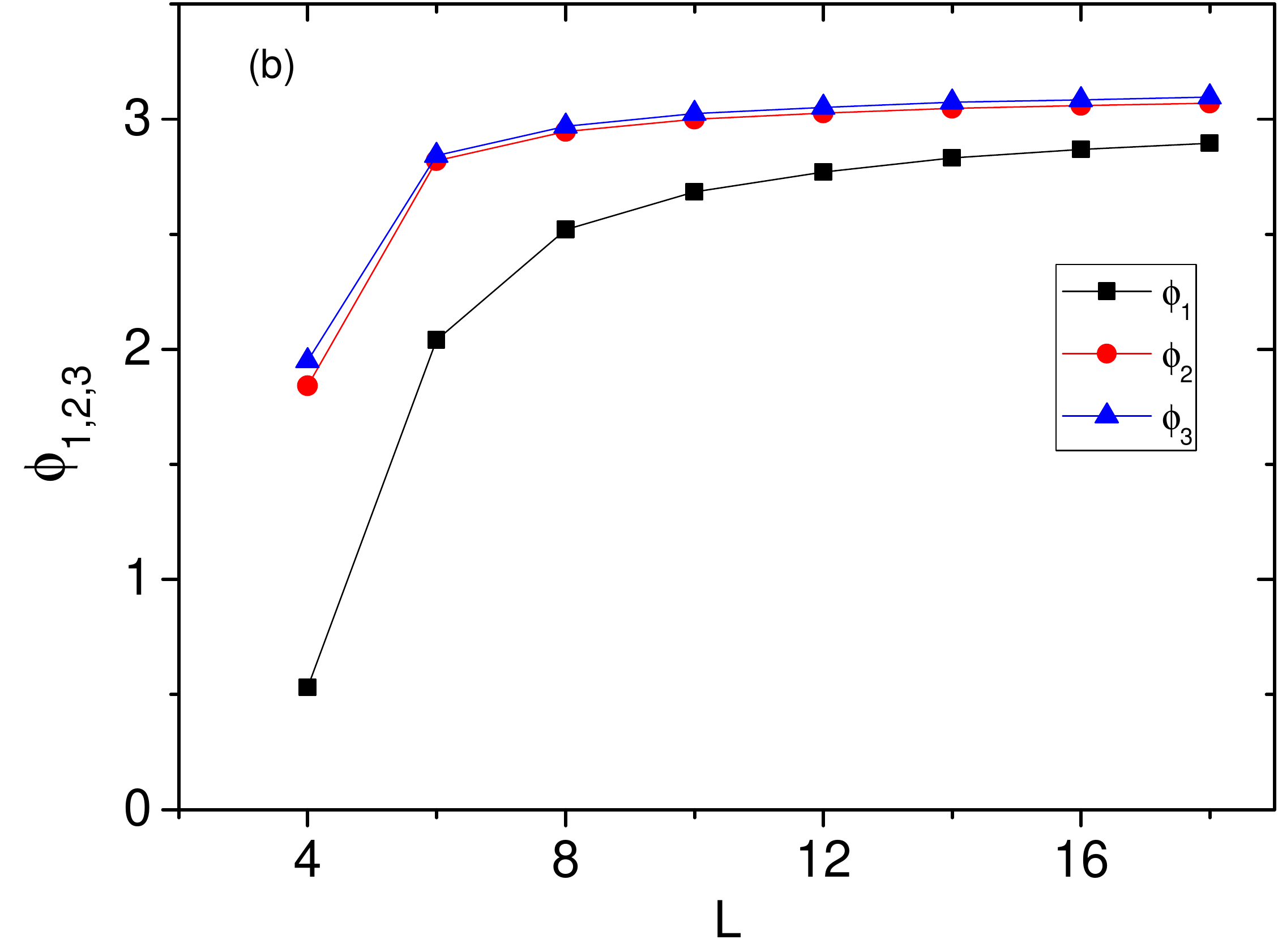}
\caption{The optimized value of the RVB parameters in the $Z_{2}$ spin liquid state, (a)the amplitudes $\mu$, $\rho$ and $\eta$, (b)the gauge angles $\phi_{1}$, $\phi_{2}$ and $\phi_{3}$. The optimized RVB parameters are found to be very close to the non-injective line $\rho=\eta$ and $\phi_{1}=\phi_{2}=\phi_{3}=\pi$, where the mapping between the mean field ansatz and the RVB state becomes singular.} \label{fig3}
\end{figure}

\section{Results and discussions}
\subsection{The variational ground state of the spin-$\frac{1}{2}$ KAFH}

The RVB parameters in the $U(1)$ and the $Z_{2}$ RVB state are optimized through variational Monte Carlo simulation.
The calculation is done on a $L\times L\times 3$ cluster with periodic - anti-periodic boundary condition. In previous studies\cite{Ran,Iqbal1,Iqbal2,Iqbal3,Iqbal4,Iqbal5}, it is claimed that the best RVB state of the spin-$\frac{1}{2}$ KAFH is described by a $U(1)$ gapless mean field ansatz. We find that this is not true. 
In Fig.2 and Fig.3, we present the optimized RVB parameters for the $U(1)$ and the $Z_{2}$ spin liquid state. The largest cluster size that we have achieved good convergence in the RVB parameters is $L=18$, beyond which the optimization procedure becomes numerically too expensive. The energy of the optimized $U(1)$ and $Z_{2}$ spin liquid state is plotted in Fig.4 as a function of $L$. The $Z_{2}$ RVB state has clearly a lower energy than the $U(1)$ state in the thermodynamic limit. 

\begin{figure}
\includegraphics[width=8cm,angle=0]{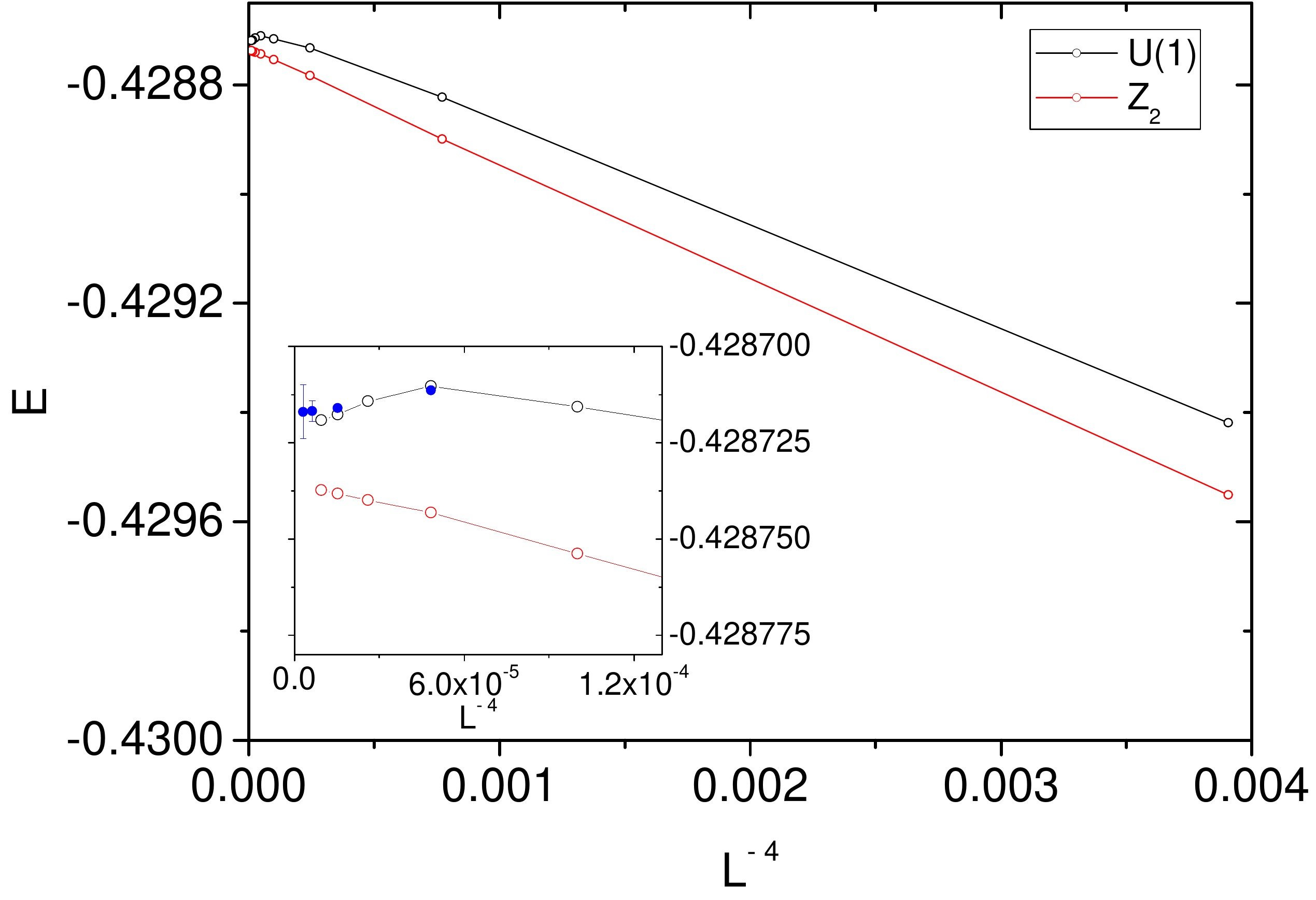}
\caption{The variational energy of the $U(1)$ and the $Z_{2}$ spin liquid state as a function of $L$. The details at large $L$ is shown in the inset. The error bar of the data is smaller than the symbol size. The blue solid circles in the inset is the result of Ref.[\onlinecite{Iqbal5}] for the $U(1)$ spin liquid state.}
\end{figure}

 \begin{figure}
\includegraphics[width=6cm,angle=0]{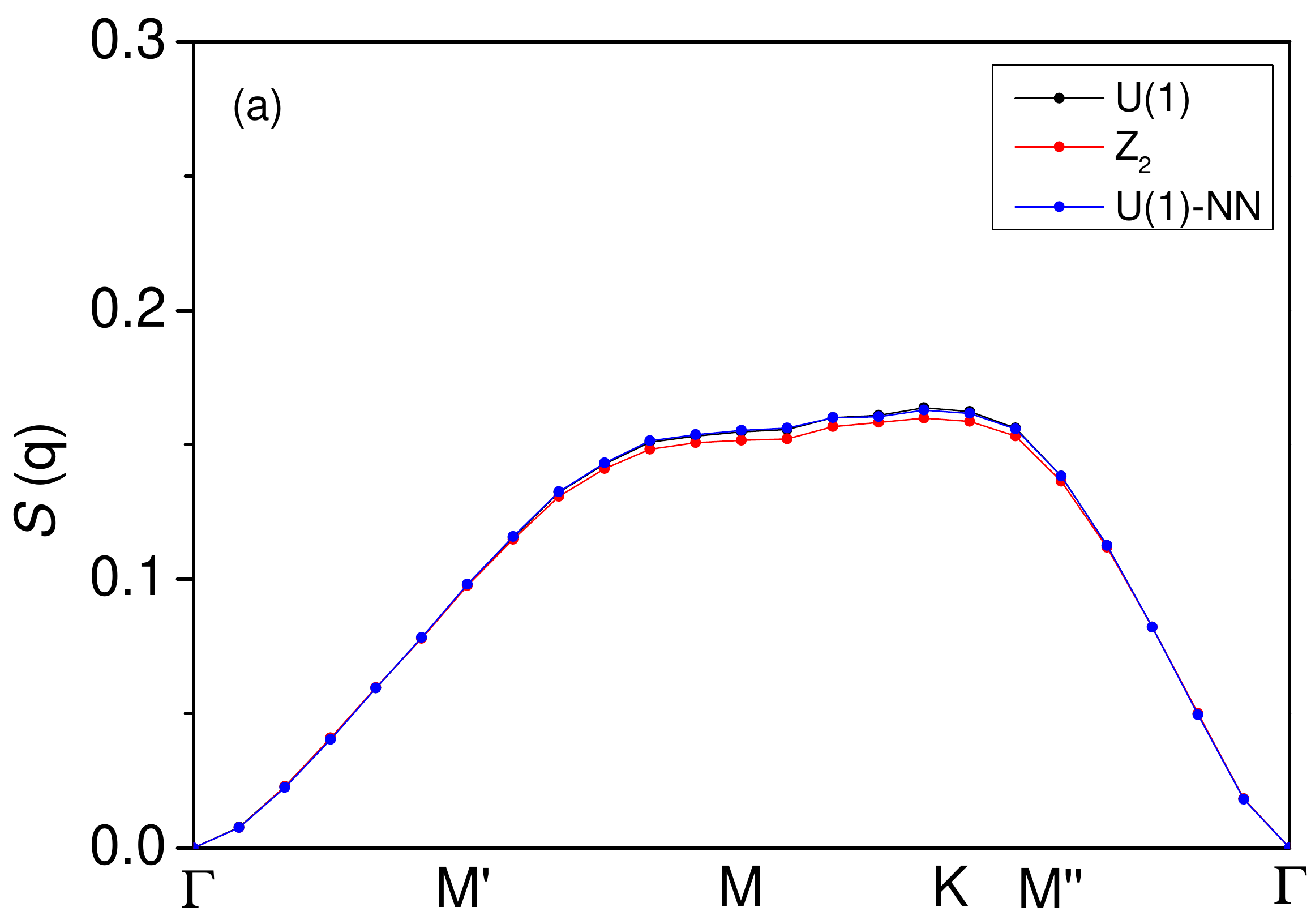}
\includegraphics[width=6cm,angle=0]{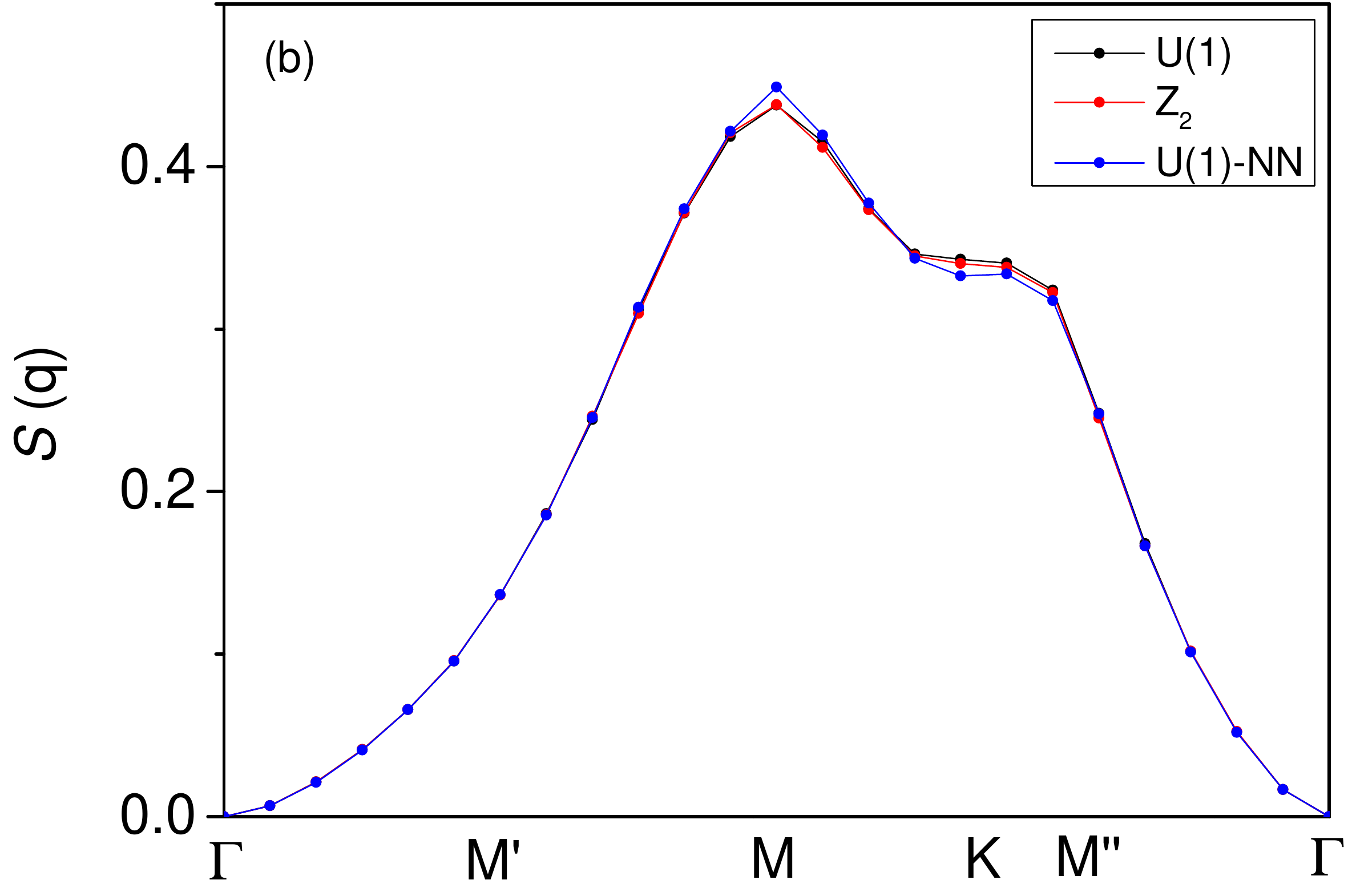}
\caption{The static spin structure factor of the $U(1)$, $Z_{2}$ and the $U(1)$-NN state along $\Gamma-\mathbf{M}'-\mathbf{M}-\mathbf{K}-\mathbf{M}''-\Gamma$. (a)Results in the mean field RVB state, (b)Results in Gutzwiller projected RVB state.}
\end{figure}  

We note that the energy difference between the $U(1)$ and the $Z_{2}$ spin liquid state is extremely small. To find out how close the two states are in the Hilbert space, we have calculated their overlap on finite clusters. We find that the overlap is still as large as $0.93$ on a $L=18$ cluster.  Both states are also very close to the $U(1)$-NN state studied in Ref.[\onlinecite{Ran}], since the optimized RVB parameters of the two states are both very close to the non-injective line. The closeness of the $U(1)$, $Z_{2}$ and the $U(1)$-NN state can be made more transparent by comparing their static spin structure factor. As is shown in Fig.6, the static spin structure factor of these states are almost identical both before and after the Gutzwiller projection. We note that the maximum of the static spin structure factor moves from the $\mathbf{K}$ point to the $\mathbf{M}$ point after the Gutzwiller projection. The strength of the spin fluctuation is also enhanced by the Gutzwiller projection.

\begin{figure}
\includegraphics[width=6cm,angle=0]{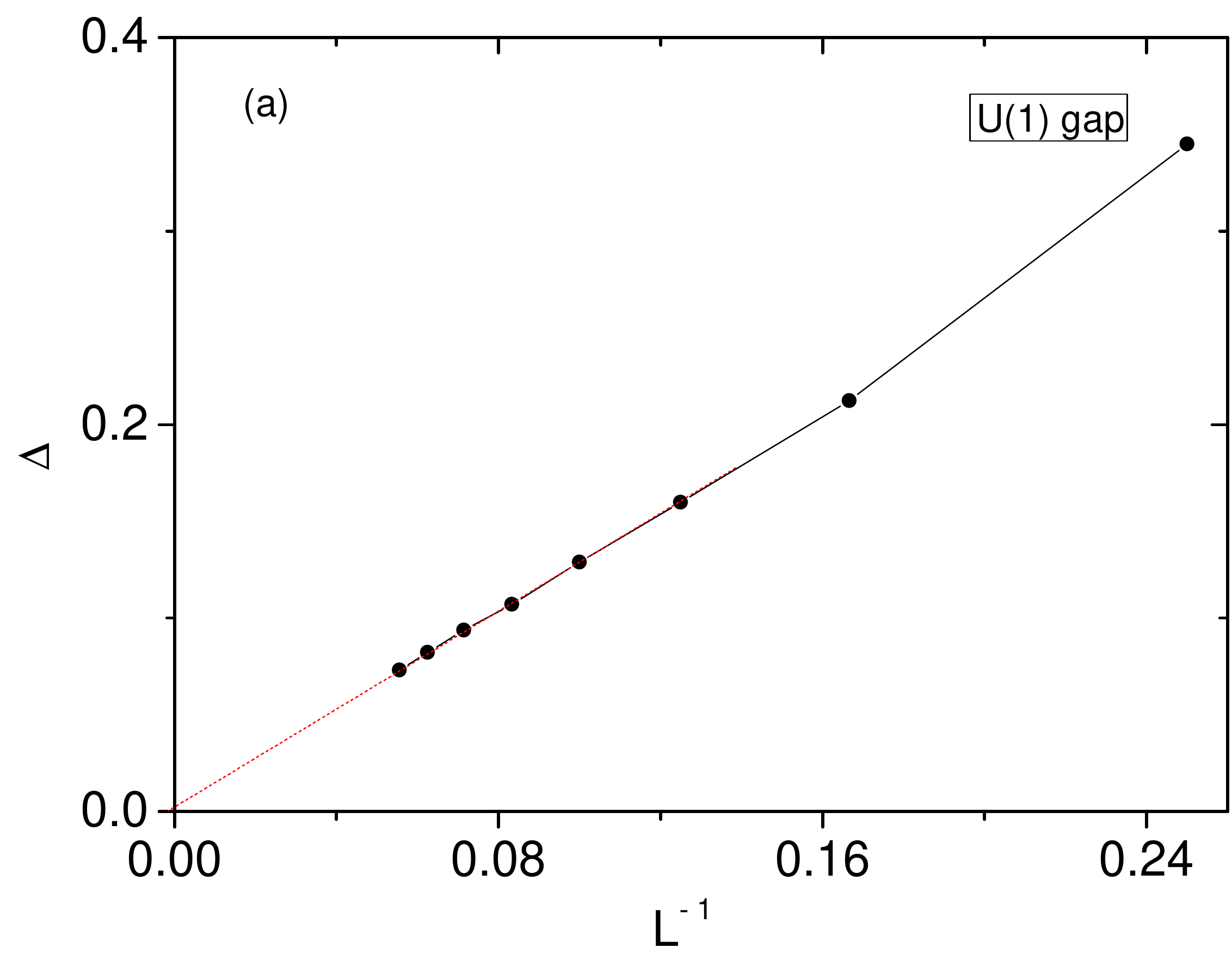}
\includegraphics[width=6cm,angle=0]{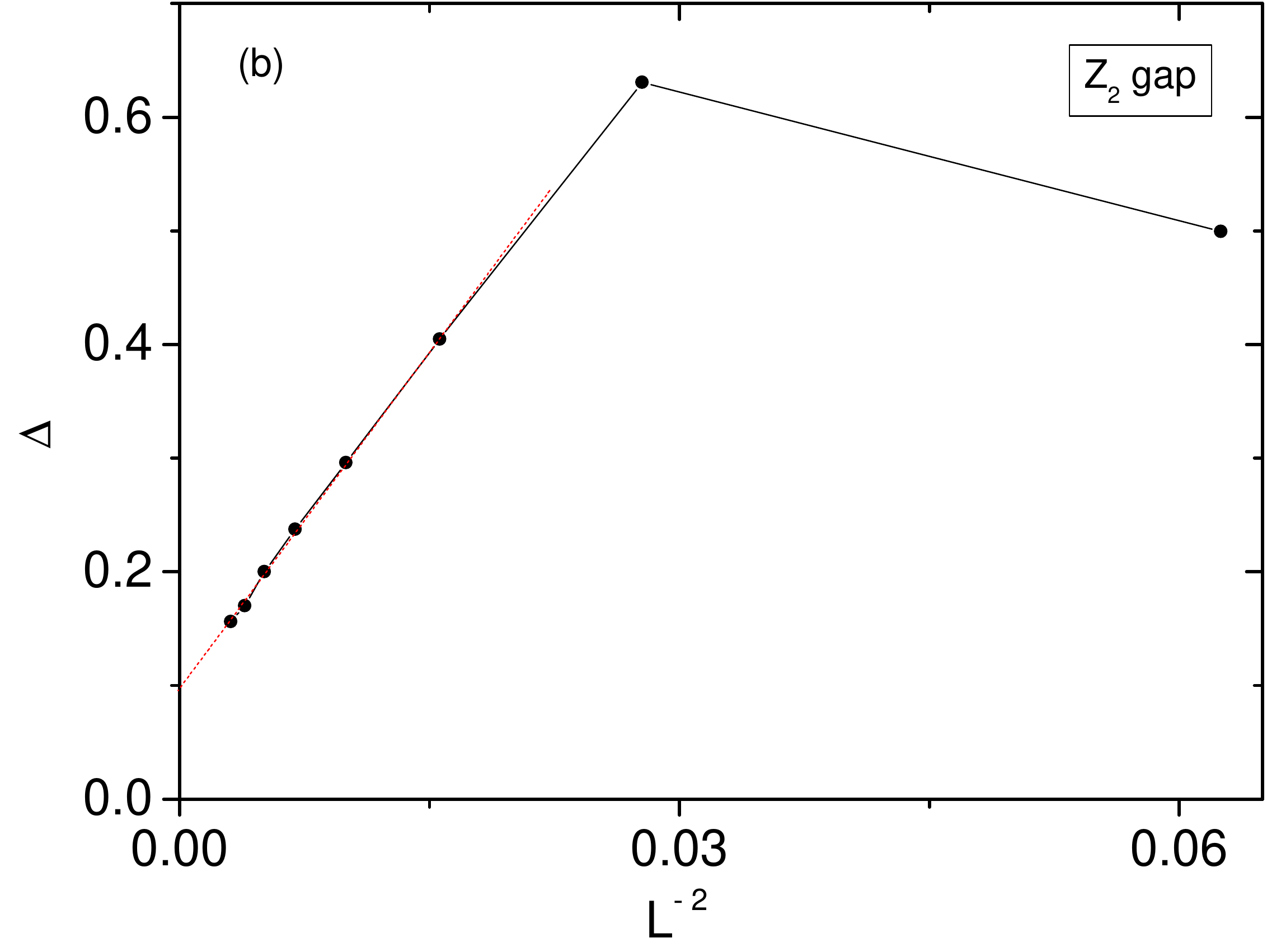}
\caption{The scaling behavior of the mean field spinon gap. The red dashed lines denote the linear fitting of the data. Note that the mean field spinon gap in the $U(1)$ and the $Z_{2}$ spin liquid state exhibit different scaling behavior at large $L$.}
\end{figure}

We now show that the $Z_{2}$ spin liquid state has indeed a finite spinon gap at the mean field level. Here we will use the RVB parameter on the nearest-neighboring bond as the unit of energy. The mean field spinon gap on finite clusters for both the $U(1)$ and the $Z_{2}$ spin liquid state are plotted in Fig.7. We find that the mean field spinon gap in the $U(1)$ spin liquid state extrapolates to zero in the thermodynamic limit as $\Delta\approx \alpha L^{-1}$, a scaling behavior naturally expected for a Dirac spin liquid. On the other hand, the mean field spinon gap of the $Z_{2}$ spin liquid state is found to approach a finite value in the thermodynamic limit as $\Delta\approx \Delta_{0}+\beta L^{-2}$, with $\Delta_{0}\simeq0.1$. Such a scaling behavior is just what one expect for a gapped system\cite{Sorella}. 

However, we note that the mean field spinon dispersion is subjected to the ambiguity related to the peculiar flat band physics of the Kagome lattice and is thus unphysical. In the next subsection, we will present the spin fluctuation spectrum calculated from the GRPA theory on the RVB state. To one's surprise, one find that the spin fluctuation spectrum above the $Z_{2}$ RVB state is actually gapless and is almost identical to that above the $U(1)$ and the $U(1)$-NN spin liquid state.

\subsection{The variational spin fluctuation spectrum of the spin-$\frac{1}{2}$ KAFH}
While the $U(1)$,  $Z_{2}$ and the $U(1)$-NN RVB state are very close to each other in the Hilbert space, they host very different mean field spinon dispersion. In Fig.8, we plot the spin fluctuation spectrum of these states calculated at the mean field level along the path $\Gamma-\mathbf{M}'-\mathbf{M}-\mathbf{K}-\mathbf{M}''-\Gamma$. The distribution of spectral weight in energy is found to be very different in these states, although the integration of spectral weight over energy, namely, the static spin structure factor $S(\mathbf{q})$, are almost identical in these states. The RVB mean field theory thus fails to provide a reliable prediction on the spin fluctuation spectrum of the spin-$\frac{1}{2}$ KAFH, if we insist on relating the excitation characteristic of a quantum system to its ground state structure.
 \begin{figure}
 \includegraphics[width=4.2cm,height=3cm]{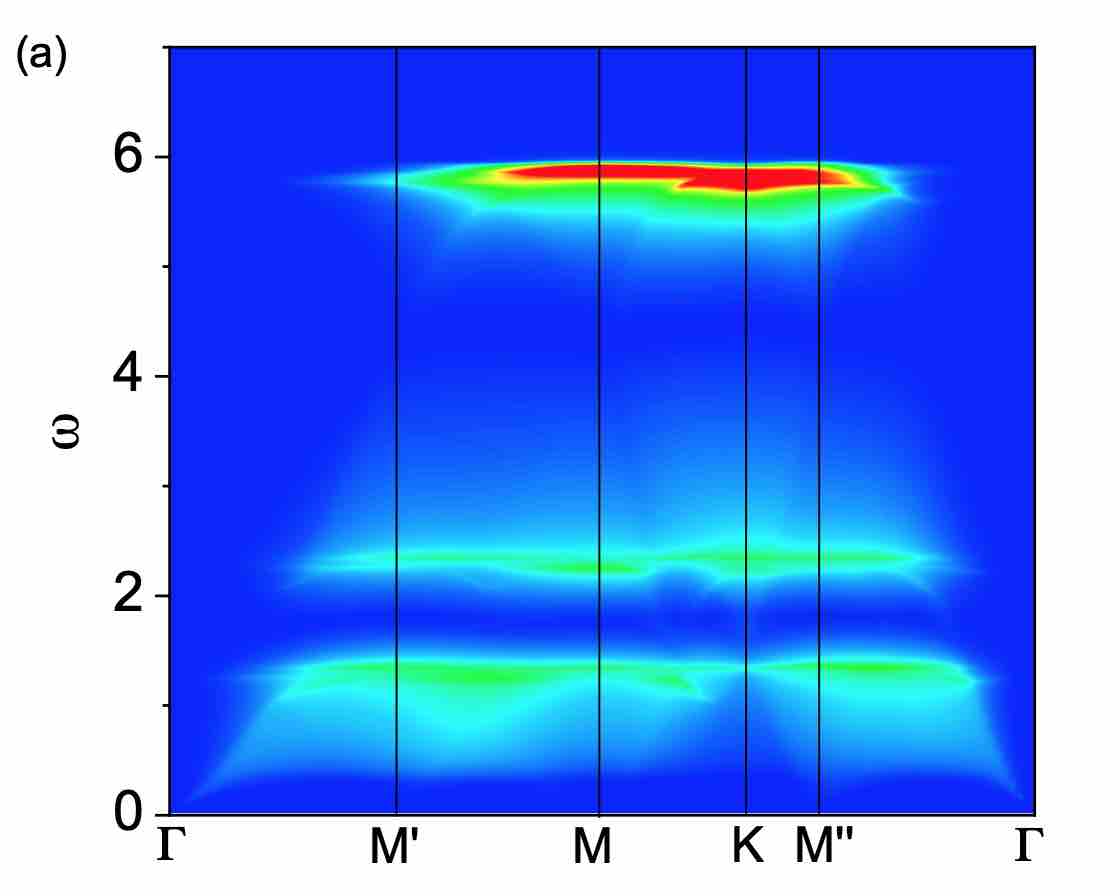}
\includegraphics[width=4.2cm,height=3cm]{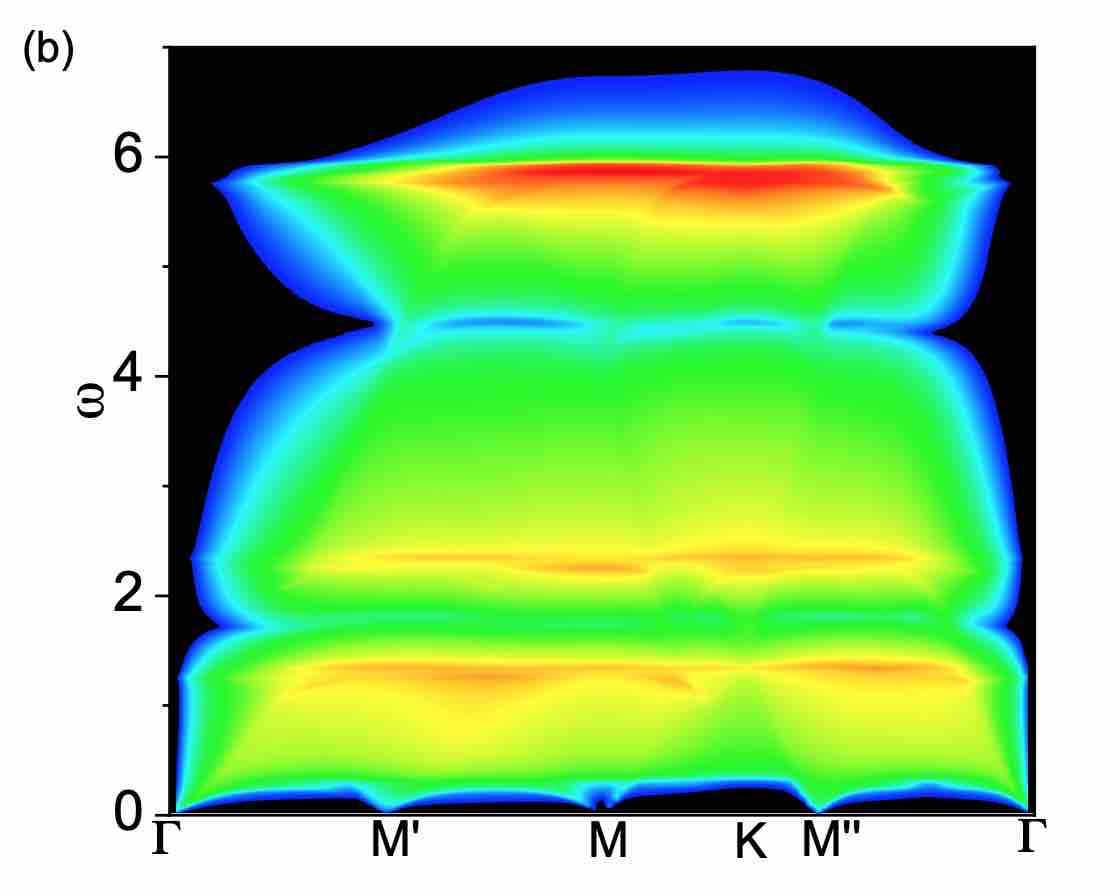}
\includegraphics[width=4.2cm,height=3cm]{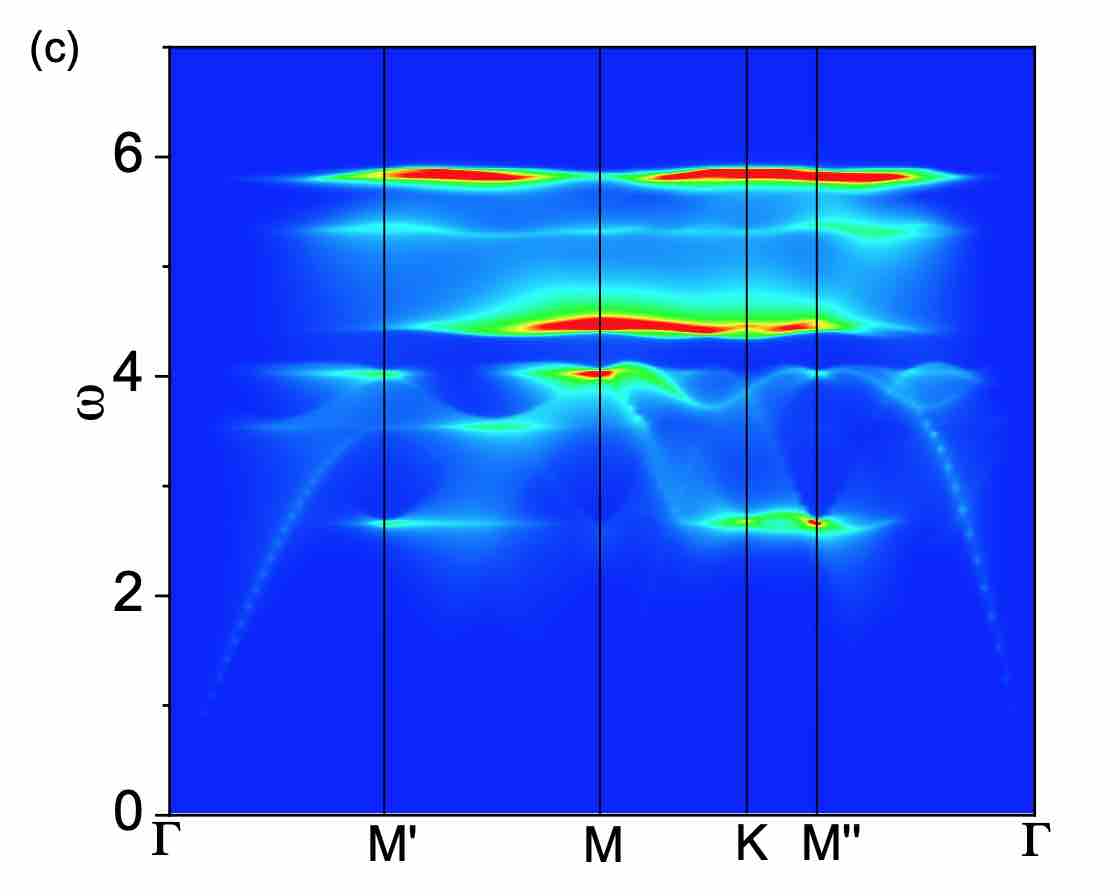}
\includegraphics[width=4.2cm,height=3cm]{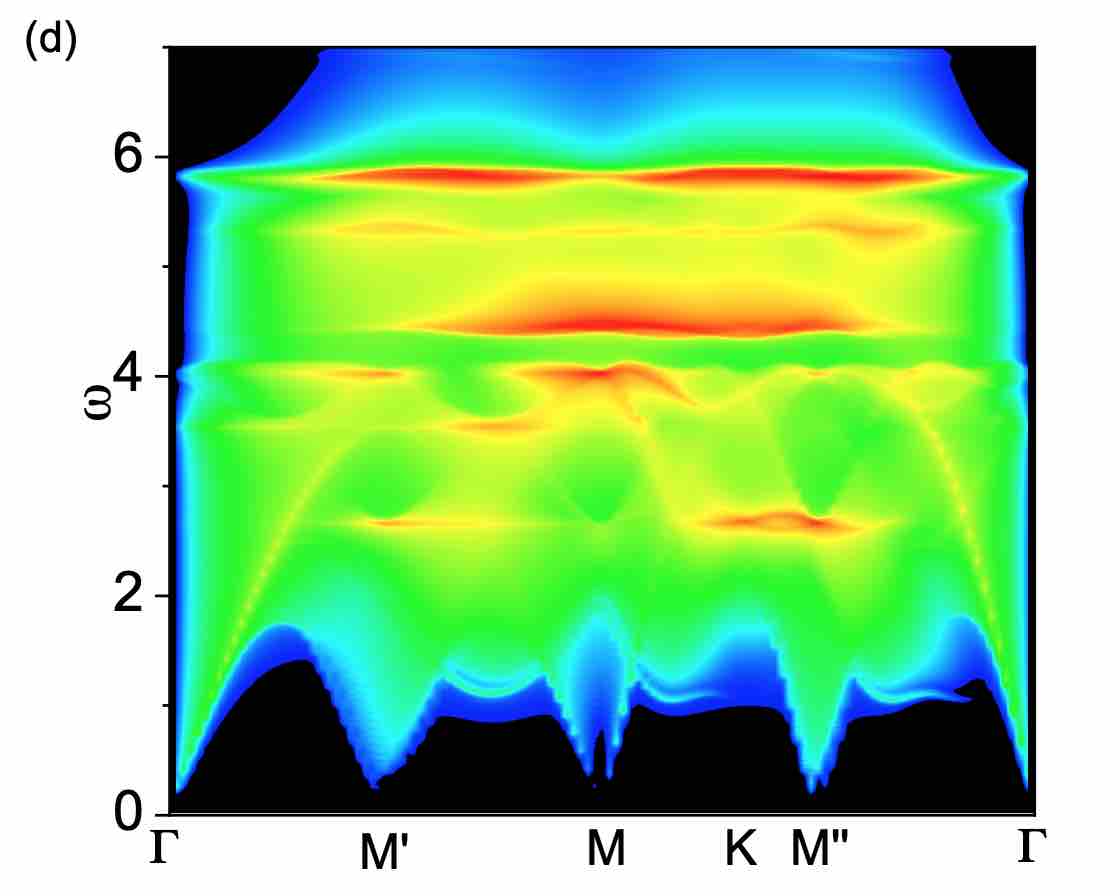}
\includegraphics[width=4.2cm,height=3cm]{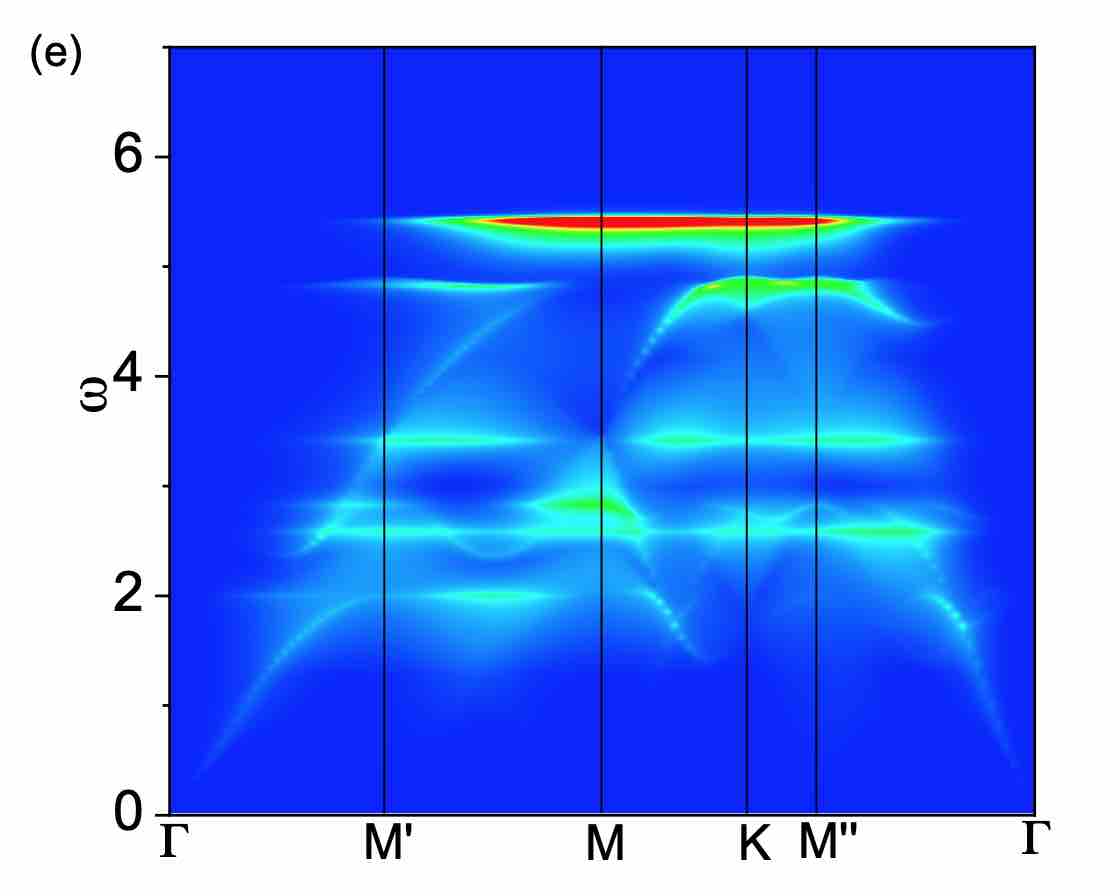}
\includegraphics[width=4.2cm,height=3cm]{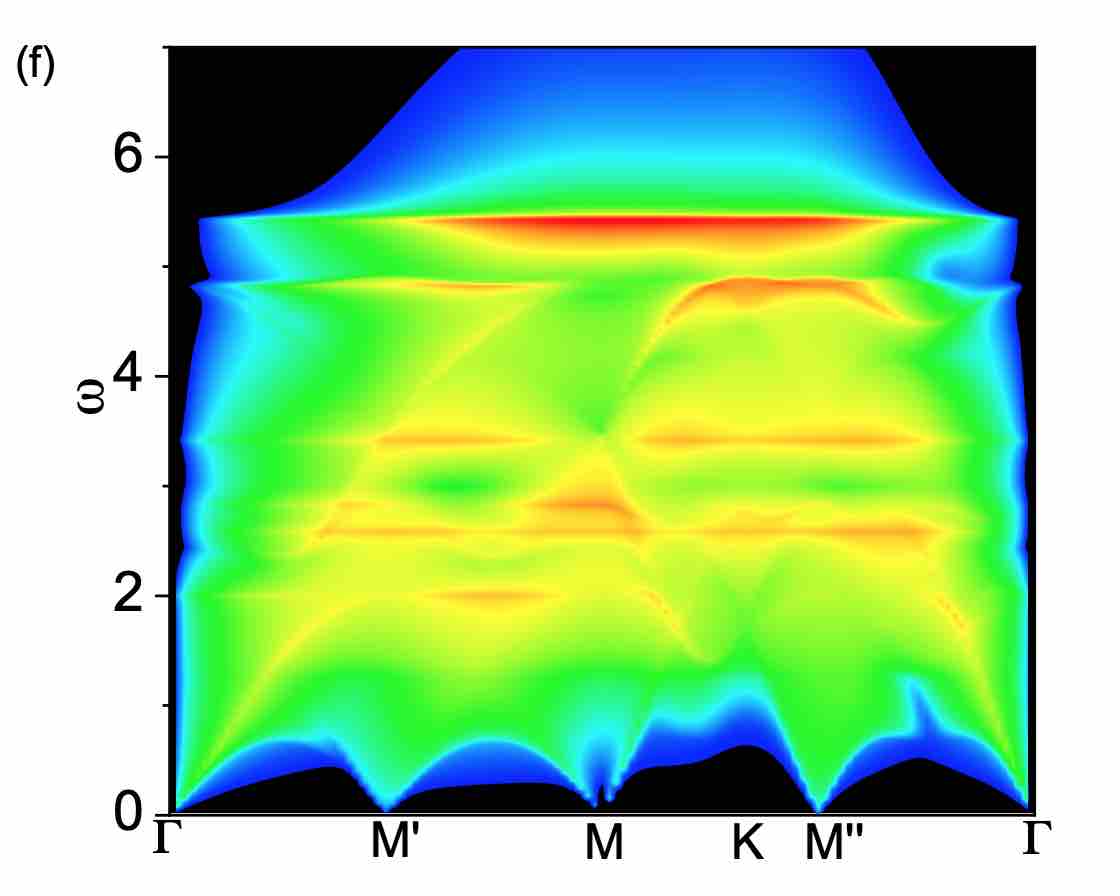}
\caption{The mean field spin fluctuation spectrum of the $U(1)$(a,b), $Z_{2}$(c,d) and the $U(1)$-NN(e,f) RVB state along $\Gamma-\mathbf{M}'-\mathbf{M}-\mathbf{K}-\mathbf{M}''-\Gamma$, plotted in linear(left column) and logarithmic(right column) scale.  The RVB parameter on the nearest neighboring bond is used as the unit of energy. The mean field spin fluctuation spectrum of the three states are found to be very different, although they have almost identical static spin structure factor(see Fig.6a).}
\end{figure}

\begin{figure}
\includegraphics[width=4.2cm,height=3cm]{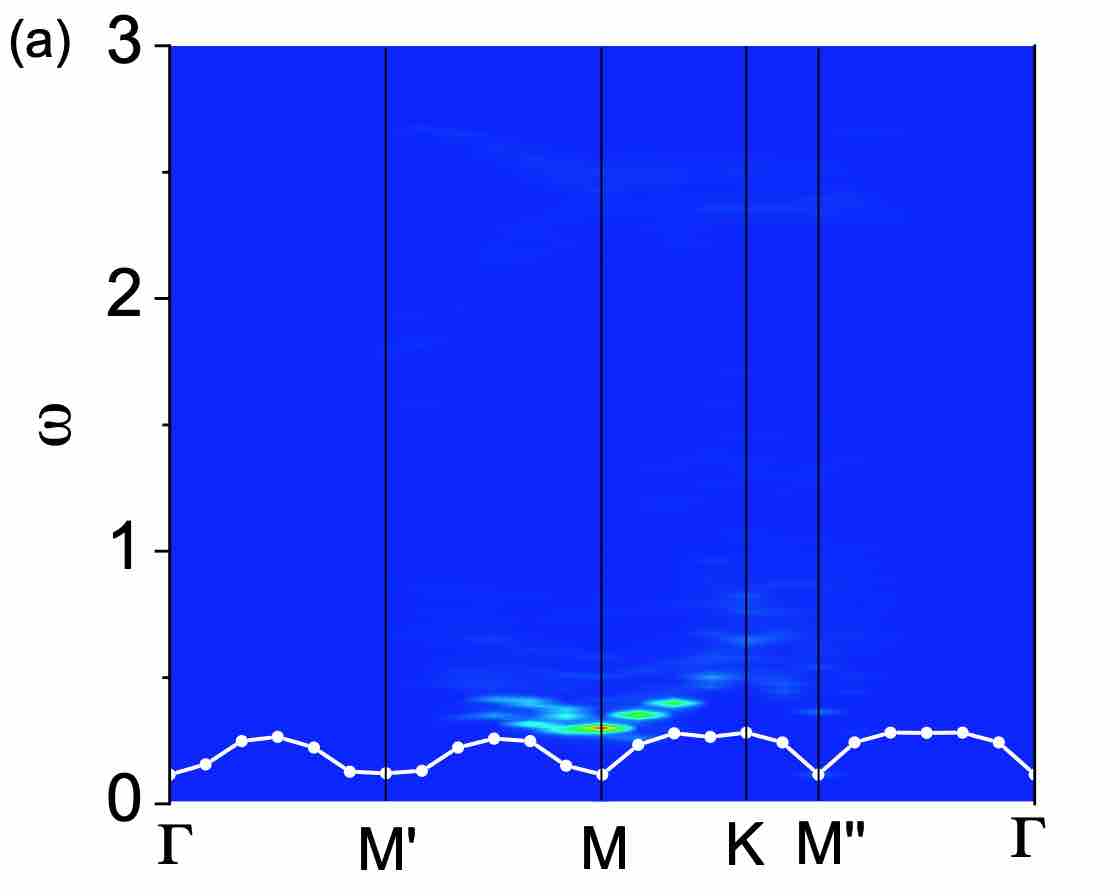}
\includegraphics[width=4.2cm,height=3cm]{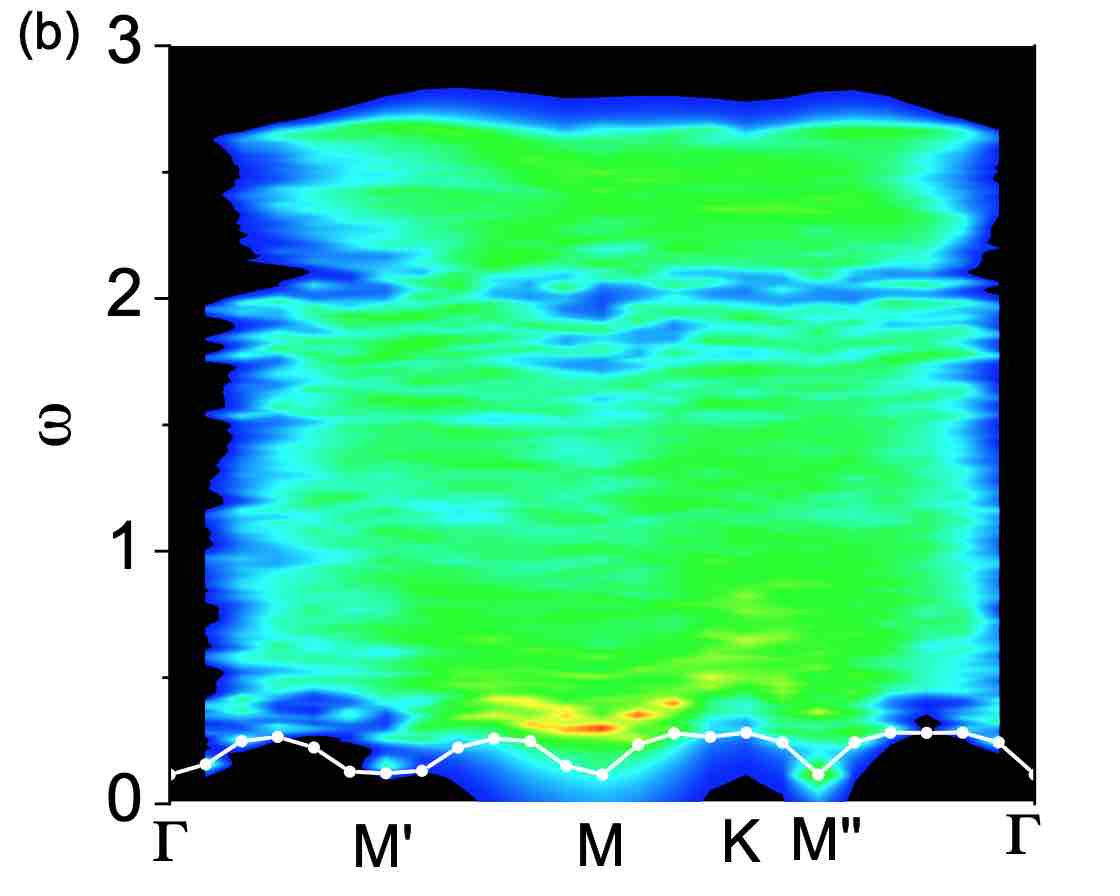}
\includegraphics[width=4.2cm,height=3cm]{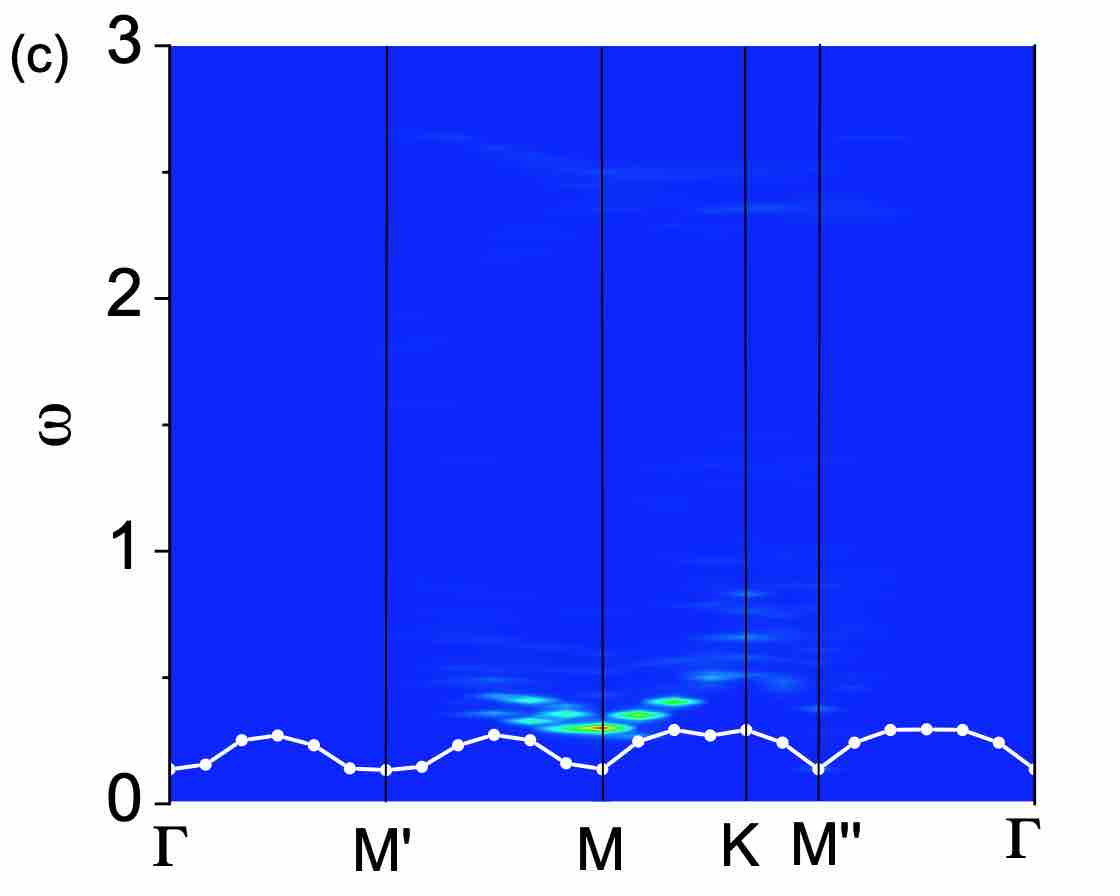}
\includegraphics[width=4.2cm,height=3cm]{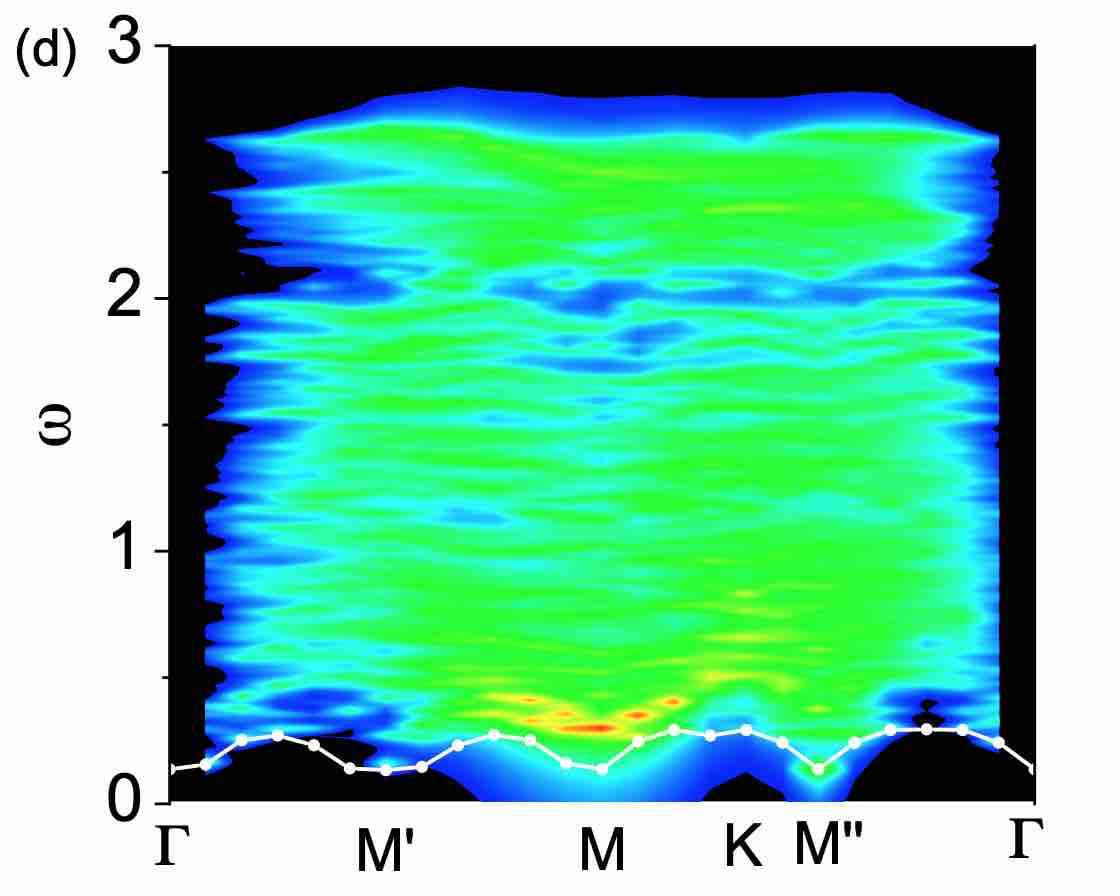}
\includegraphics[width=4.2cm,height=3cm]{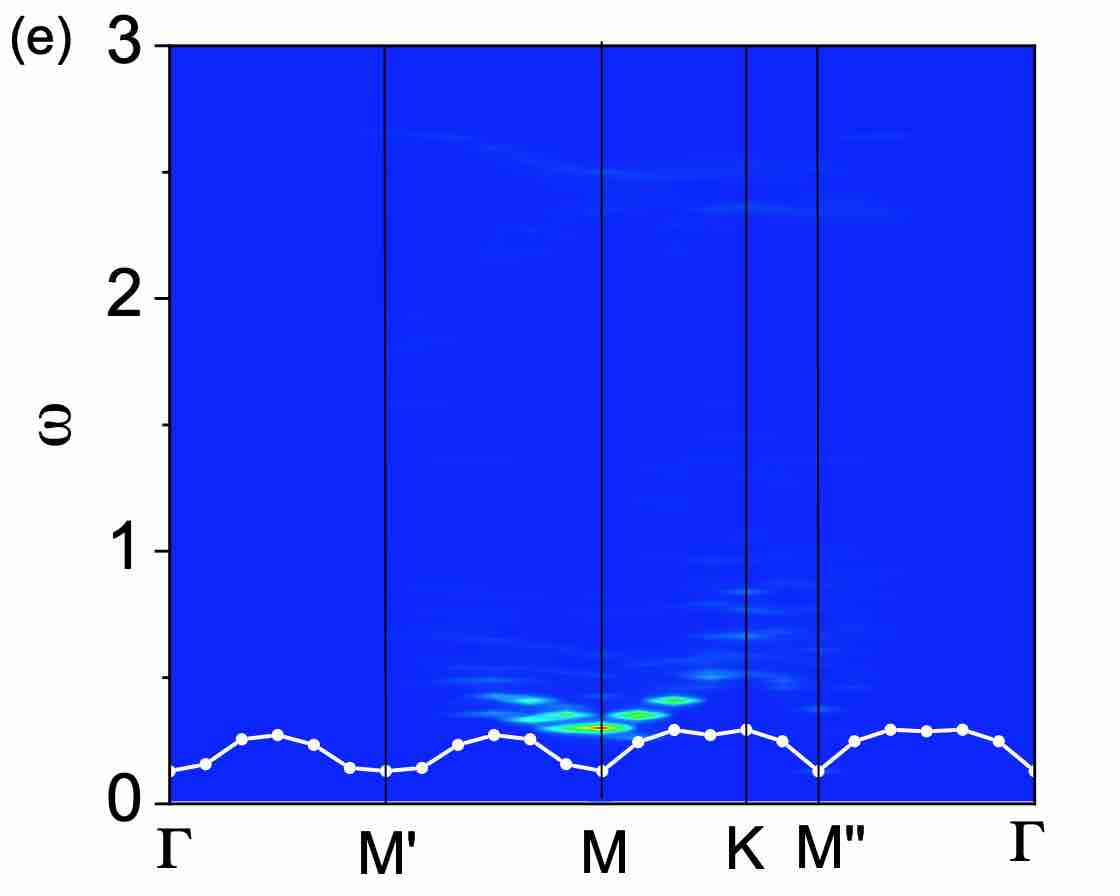}
\includegraphics[width=4.2cm,height=3cm]{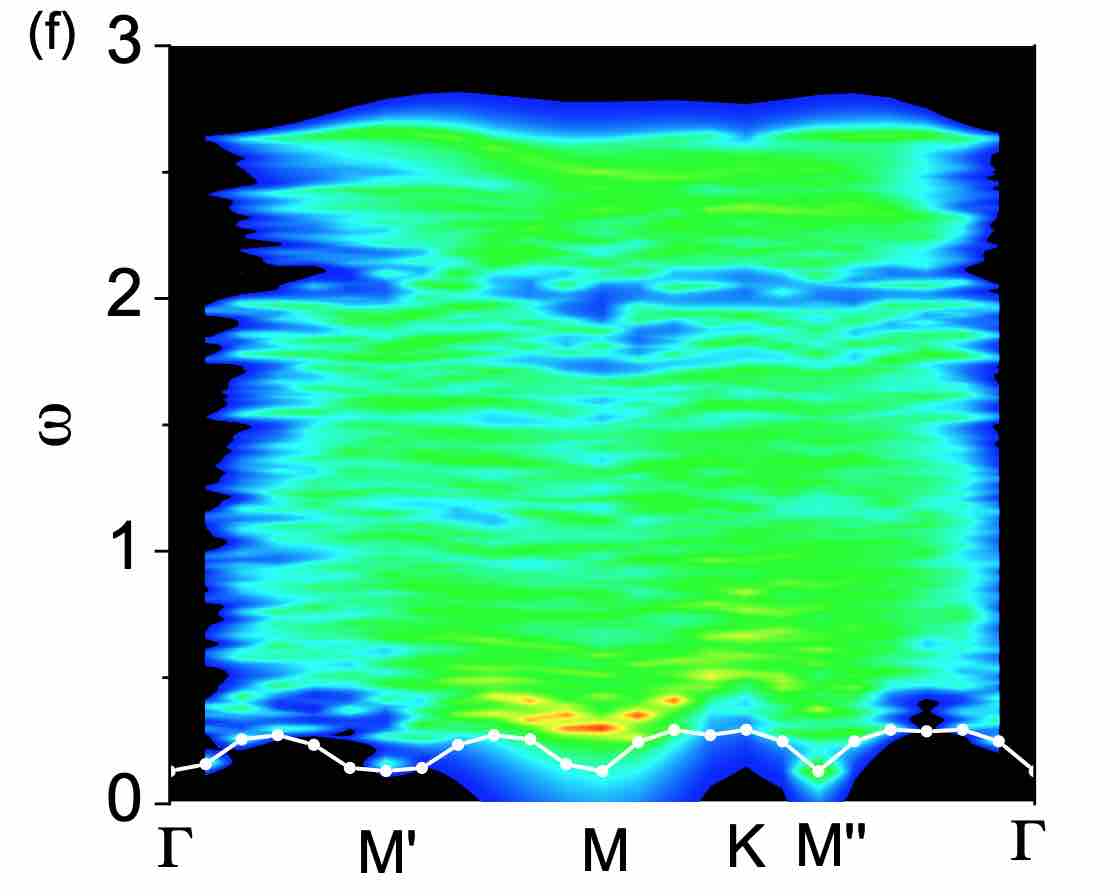}
\includegraphics[width=7cm,angle=0]{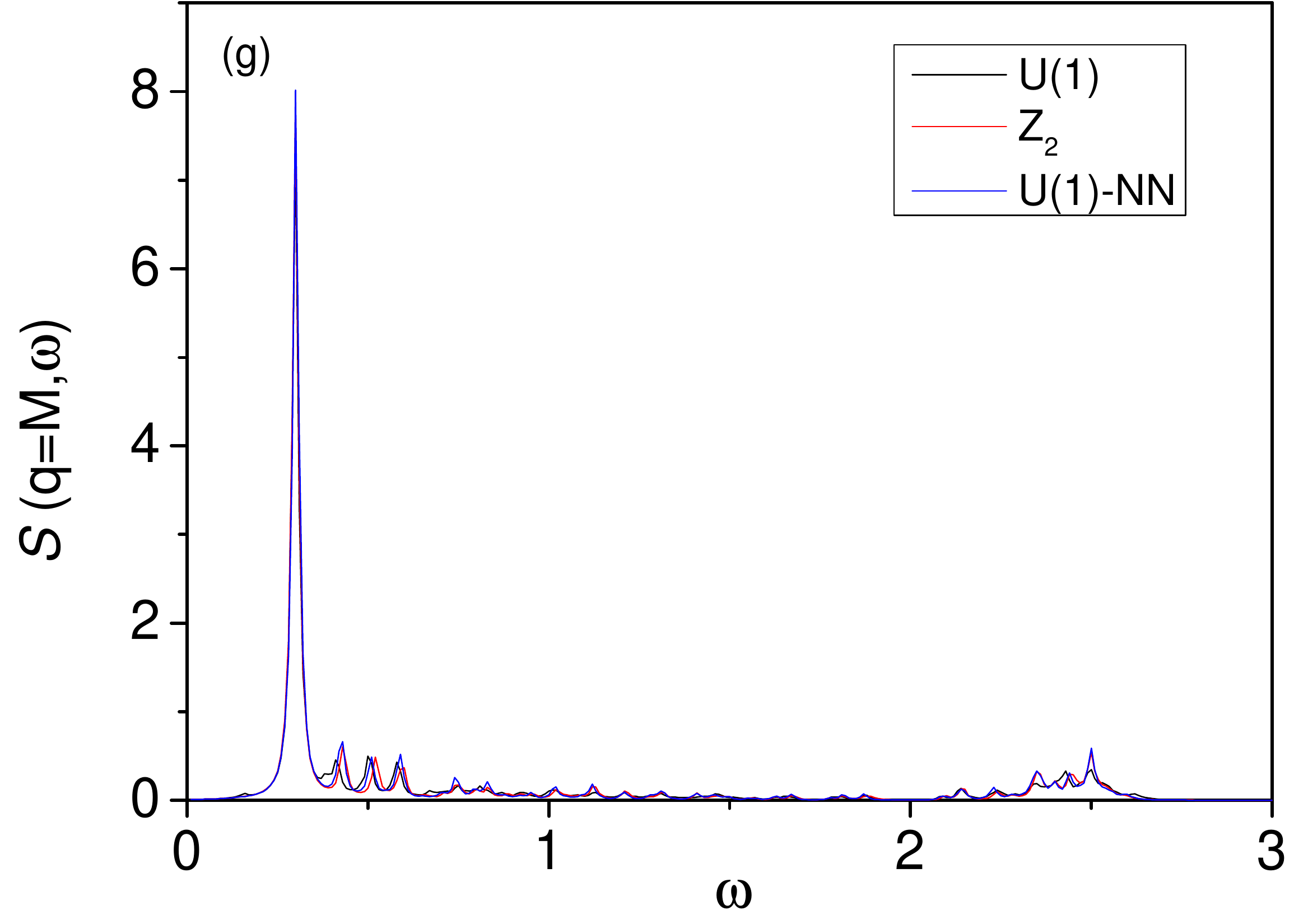}
\caption{The spin fluctuation spectrum calculated from the GRPA theory for the $U(1)$(a,b), $Z_{2}$(c,d) and the $U(1)$-NN(e,f) RVB state along $\Gamma-\mathbf{M}'-\mathbf{M}-\mathbf{K}-\mathbf{M}''-\Gamma$, plotted in linear(left column) and logarithmic(right column) scale. The Heisenberg exchange coupling $J$ is used as the unit of energy. The spin fluctuation spectrum above the three states are found to be almost identical, as can be seen more clearly in (g), in which we compare the spin fluctuation spectrum of the three states at the $\mathbf{M}$ point. The white lines in (a)-(f) mark the lower boundary of the spin fluctuation continuum.}
\end{figure}  

We now go beyond the mean field theory. The spin fluctuation spectrum calculated from the GRPA theory is plotted in Fig.9 for the $U(1)$, $Z_{2}$ and the $U(1)$-NN RVB state. The calculation is done on a $L=12$ cluster with a broadening of $\delta=0.01J$ in energy. To one's surprise, the spin fluctuation spectrum above all these three states are found to be almost identical with each other. This is, however, just what one should expect if we note that these three states are very close to each other in the Hilbert space. Equally surprising is the huge difference between the spin fluctuation spectrum calculated from the GRPA theory and that from the RVB mean field theory. More specifically, the spin fluctuation spectrum calculated from the GRPA theory is characterized by a prominent spectral peak around the M point, with an energy of about $0.3J$ for $L=12$. The remaining spin fluctuation spectral weight is distributed in a broad and almost featureless continuum extending to an energy as high as $2.7J$\cite{note}. 

\begin{figure}
\includegraphics[width=6cm,angle=0]{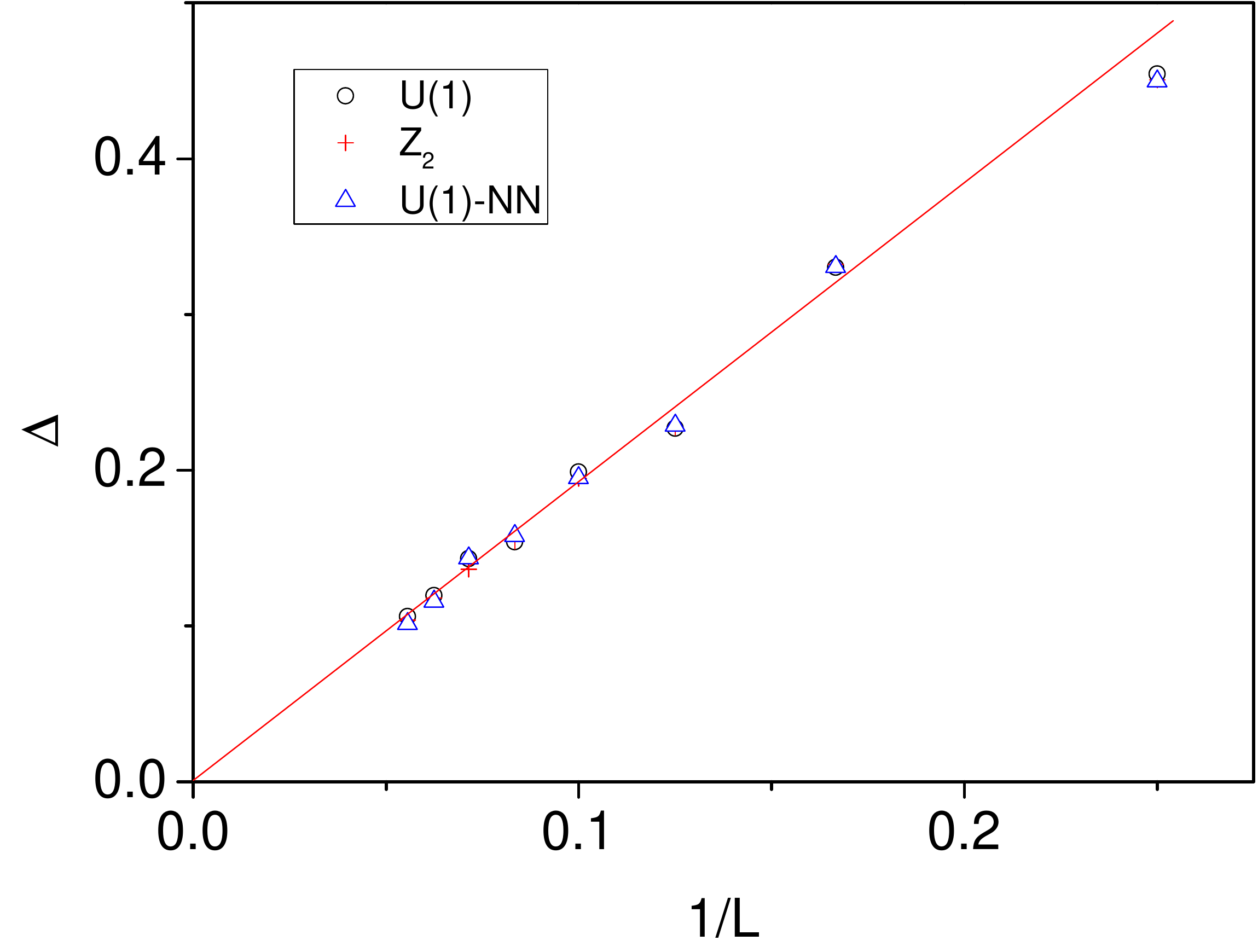}
\caption{Scaling of the spin gap in the $U(1)$, $Z_{2}$ and $U(1)$-NN RVB state with $1/L$. The Heisenberg exchange coupling $J$ is used as the unit of energy.}
\end{figure}  

\begin{figure}
\includegraphics[width=6cm,angle=0]{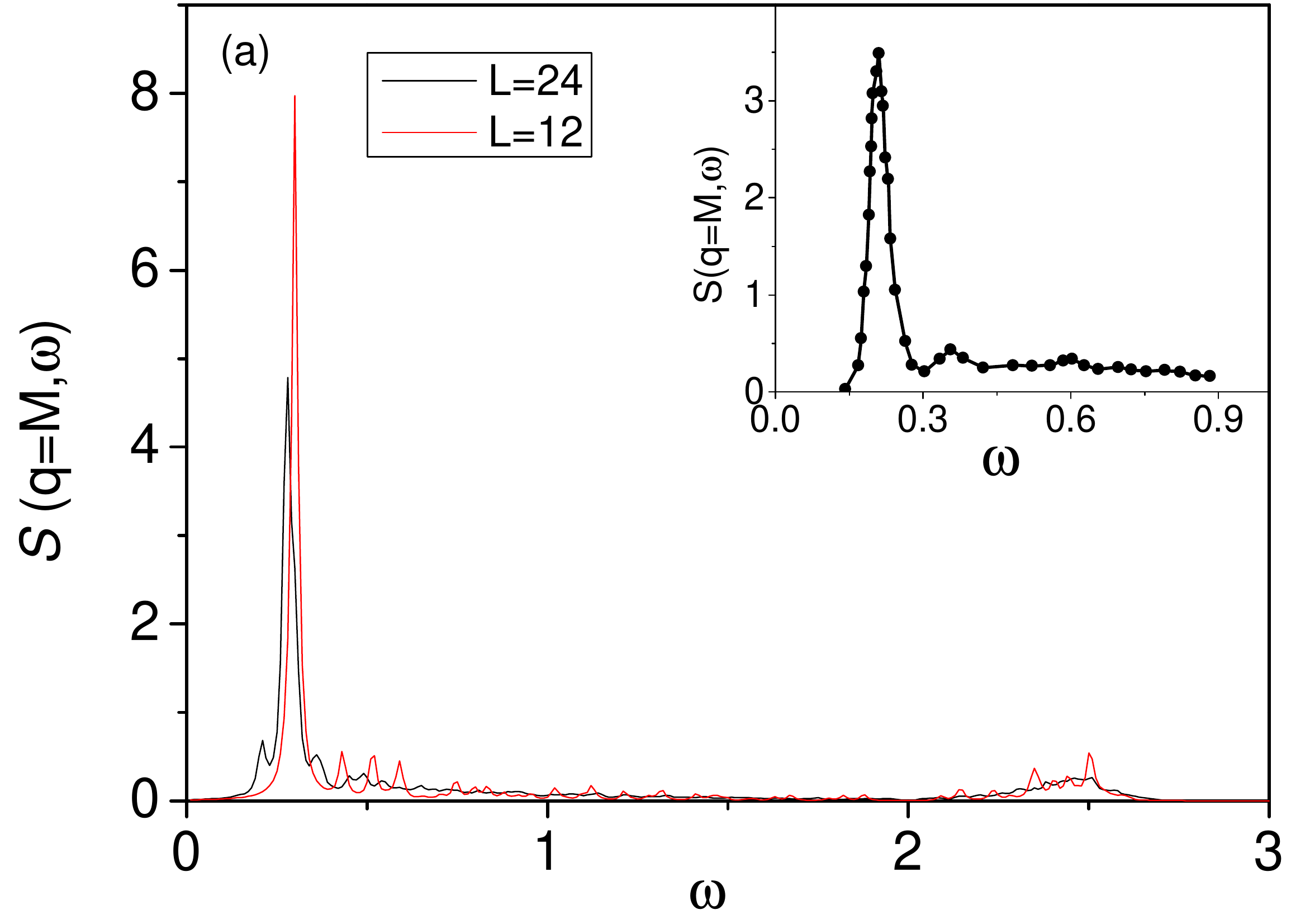}
\includegraphics[width=6cm,angle=0]{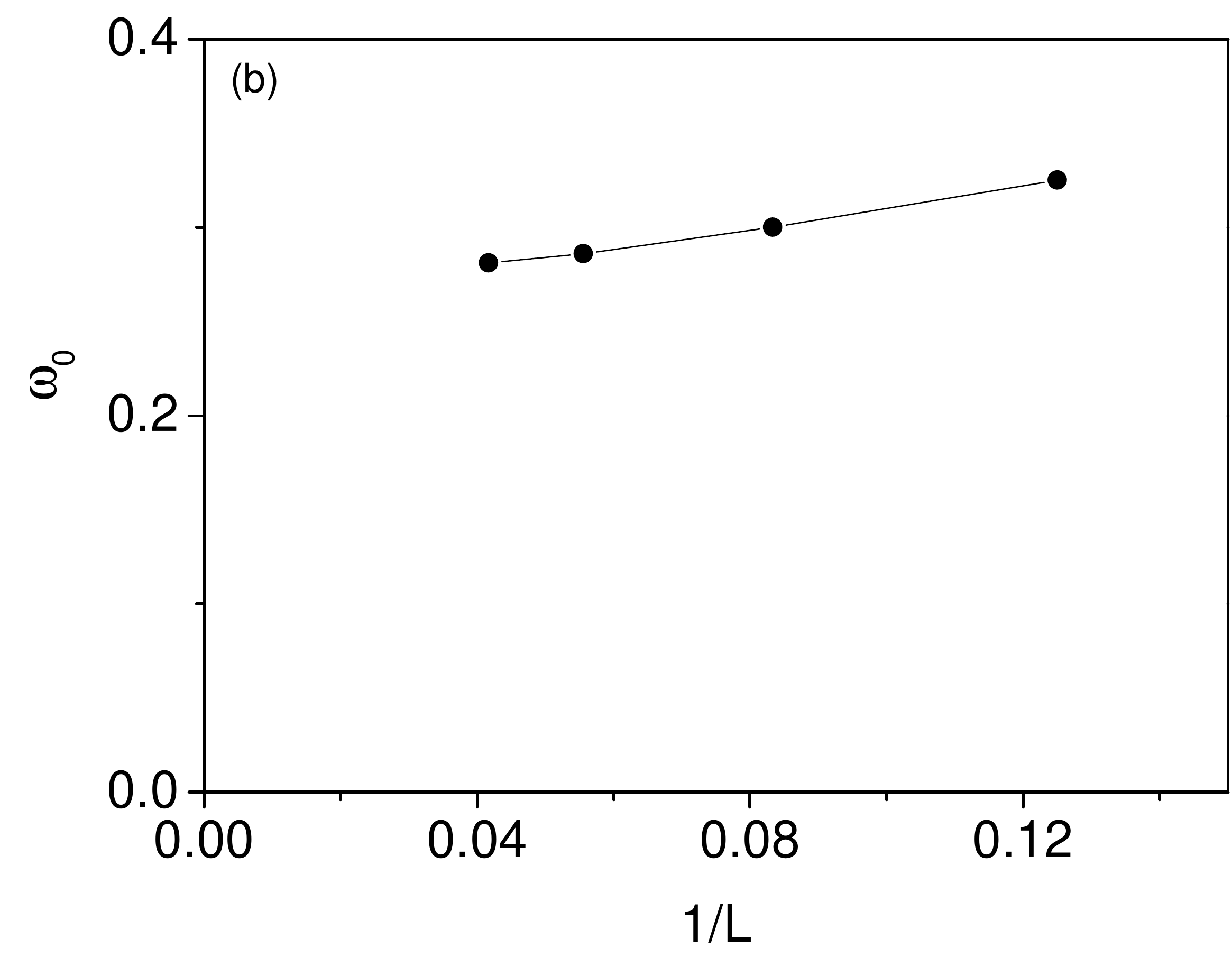}
\caption{(a)The spin fluctuation spectrum at the $\mathbf{M}$ point for $L=12$ and $L=24$. The Heisenberg exchange coupling $J$ is used as the unit of energy. Shown in the inset is the DMRG result reproduced from Ref.[\onlinecite{Zhu}]. (b)The energy of the intense spectral peak at the $\mathbf{M}$ point, here denoted as $\omega_{0}$, as a function of $1/L$.}
\end{figure}  

We note that the intense spectral peak around the M point lies within the spin excitation continuum, whose lower boundary is marked by the white lines in Fig.9. The lower boundary of the continuum reaches its minimum at the $\bm{\Gamma}$, $\mathbf{M}$, $\mathbf{M}'$ and the $\mathbf{M}''$ point(as is predicted correctly by the RVB mean field theory) and defines the spin gap of the system. We find that the spin gap so defined decreases with $L$ and extrapolates to zero in the thermodynamical limit for all these three RVB states, as can be seen in Fig.10. The $Z_{2}$ RVB state we found is thus actually a gapless spin liquid state, although it is described by a gapped mean field ansatz.

To see more clearly the spectral weight distribution in energy, we plot in Fig.11 the spin fluctuation spectrum at the $\mathbf{M}$ point of the $L=12$ and $L=24$ cluster. Here we only present the result of the $U(1)$-NN state for clarity, since the spectrum of the other two states are almost identical. The most prominent feature of the spectrum at the $\mathbf{M}$ point is the intense spectral peak inside the broad continuum, which contains about half of the total spin fluctuation spectral weight. We note that the same spectral characteristic is found also in dynamical DMRG simulation of the spin-$\frac{1}{2}$ KAFH\cite{Zhu}, the result of which is shown in the inset of Fig.11(a) for comparison. The existence of such an intense spectral peak around $\mathbf{M}$ point is thus a robust feature in the spin fluctuation spectrum of the spin-$\frac{1}{2}$ KAFH.  We find that the energy of this peak, here denoted as $\omega_{0}$, decreases with the system size and extrapolates to a value of no less than $0.25J$ in the thermodynamic limit(see Fig.11(b)).

\subsection{Comparison with the experimental results on Hebertsmithite ZnCu$_{3}$(OH)$_{6}$Cl$_{2}$} 

Previous INS measurement finds that the spin fluctuation spectrum of Hebertsmithite ZnCu$_{3}$(OH)$_{6}$Cl$_{2}$, which is believed to be an ideal realization of the spin-$\frac{1}{2}$ KAFH, is characterized by a featureless continuum above 2 meV\cite{Han}. Below 2 meV, the spectral intensity increases with decreasing energy and aggregates toward the $\mathbf{M}$ point in momentum space. As a result of such a trend, a broad peak emerges around the $\mathbf{M}$ point below 2 meV.

It is widely believed that such a low energy spectral peak should be attributed to Cu$^{2+}$ impurity spins occupying the Zn$^{2+}$ site between the Kagome layers, rather than the intrinsic spin fluctuation of the Kagome layer. According to such a picture, the intrinsic spin fluctuation spectrum of the Kagome layer is characterized by a featureless continuum with probably a small gap. This is supported by a later NMR study on the system\cite{Fu}, in which the Knight shift on the Oxygen site is measured. It is found that the Knight shift vanishes in the zero temperature limit for Oxygen site far away from the Cu$^{2+}$ impurity spin. A spin gap of the order of $0.03-0.07J$ is claimed by fitting the temperature dependence of the Knight shift data on such Oxygen site.

However, there are three reasons to object such a picture. First, if the spectral peak below 2 meV is indeed contributed by the Cu$^{2+}$ impurity spins, it should not exhibit such a strong momentum dependence as observed in the INS measurement. Second, as the intense spectral peak around the $\mathbf{M}$ point is such a prominent feature in the spin fluctuation spectrum of the spin-$\frac{1}{2}$ KAFH, in particular, as it contains almost half of the total spin fluctuation spectral weight at the $\mathbf{M}$ point, it must appear somewhere in the INS spectrum of Hebertsmithite ZnCu$_{3}$(OH)$_{6}$Cl$_{2}$, if the latter is indeed an ideal realization of the spin-$\frac{1}{2}$ KAFH. It is thus very likely that the peak below 2 meV in the INS spectrum of Hebertsmithite ZnCu$_{3}$(OH)$_{6}$Cl$_{2}$ corresponds just to such a theoretically predicted spectral feature. Third, since the nuclear spin on the Oxygen site is coupled symmetrically to the two neighboring Cu$^{2+}$ spins but is almost decoupled from the third Cu$^{2+}$ spin in the Kagome unit cell, a very small Knight shift on the Oxygen site does not necessarily imply a very small uniform spin susceptibility of the system. It may simply imply that the two neighboring Cu$^{2+}$ spins that the Oxygen nuclear spin are symmetrically coupled to are antiferromagnetically correlated with each other. We note that the most recent Knight shift result on the Oxygen site of Hebertsmithite ZnCu$_{3}$(OH)$_{6}$Cl$_{2}$ is consistent with a power law(rather than exponential) decay of the uniform susceptibility with temperature, which implies that the intrinsic spin fluctuation spectrum of Kagome layer is gapless\cite{Mendels}.

\begin{figure}
\includegraphics[width=6cm,angle=0]{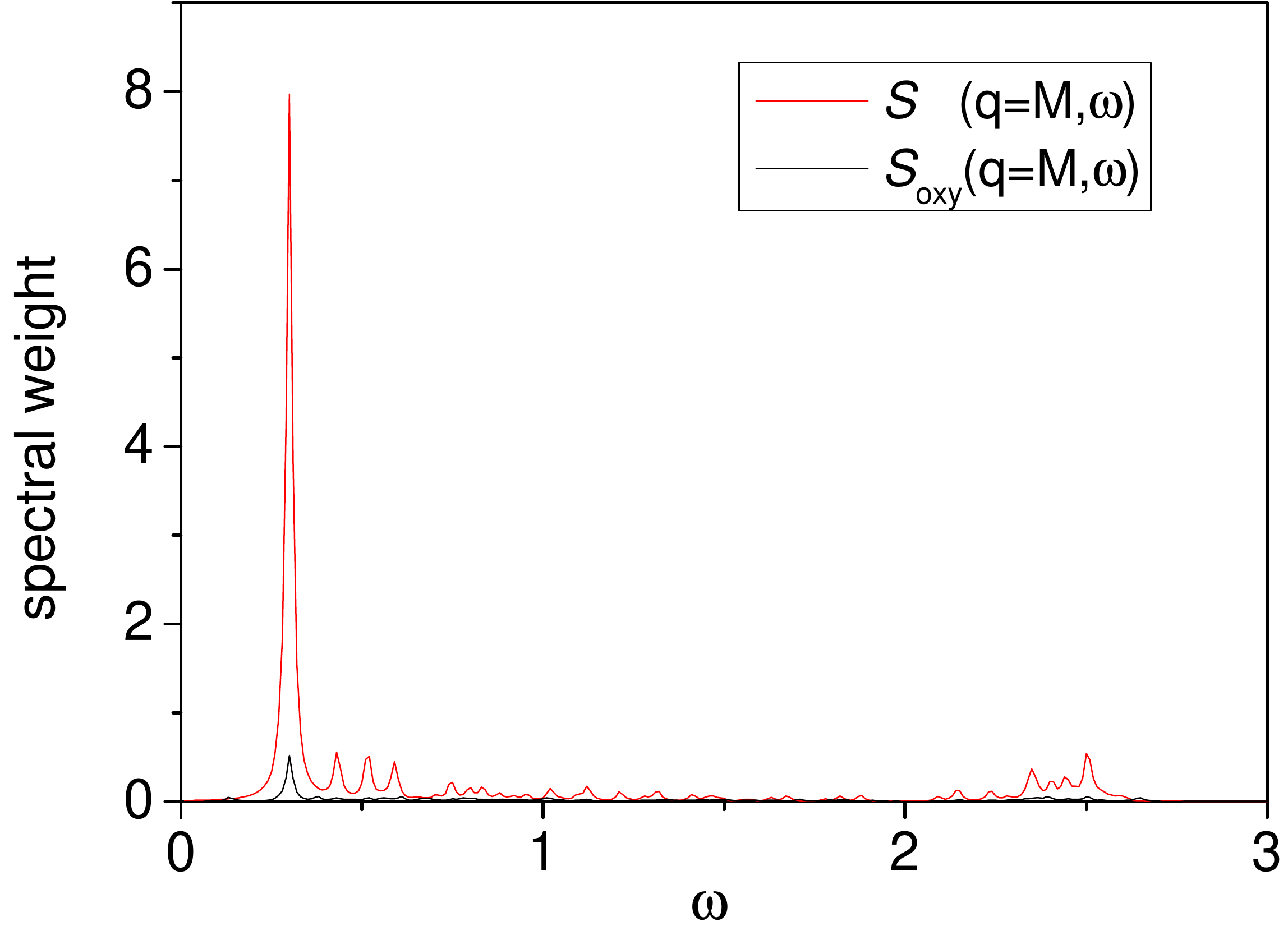}
\caption{Comparison between the fluctuation spectrum of the local field on the Oxygen site and the full spin fluctuation spectrum at the $\mathbf{M}$ point. The Heisenberg exchange coupling $J$ in the spin-$\frac{1}{2}$ NN-KAFH is used as the unit of energy. The Oxygen site is found to be almost blind to the intense spectral peak in $S(\mathbf{q=M},\omega)$.}
\end{figure}   

To verify such a picture, we have studied the fluctuation spectrum of the hyperfine field at the Oxygen site, which is proportional to the imaginary part of the following local spin susceptibility
\begin{equation}
\bm{\chi}^{z,z}_{\mathrm{oxy}}(\mathbf{q},\tau)=-\langle \ T_{\tau} \mathrm{S}^{z}_{\mathrm{oxy}}(\mathbf{q},\tau)\  \mathrm{S}^{z}_{\mathrm{oxy}}(-\mathbf{q},0) \rangle.
 \end{equation}
Here 
\begin{equation}
\mathrm{S}^{z}_{\mathrm{oxy}}(\mathbf{q})=\frac{1}{2}\sum_{\mathbf{k},\mu=1,2} \  \psi^{\dagger}_{\mathbf{k+q},\mu}\psi_{\mathbf{k},\mu},
\end{equation} 
denotes the sum of the spin density operator on sublattice 1 and 2 at momentum $\mathbf{q}$. In Fig.12, we compare the spectral function of such fluctuation, here denoted as $S_{\mathrm{oxy}}(\mathbf{q},\omega)$, with the full spin fluctuation spectrum $S(\mathbf{q},\omega)$ at $\mathbf{q=M}$. It is found that $S_{\mathrm{oxy}}(\mathbf{q=M},\omega)$ is almost blind to the intense spectral peak in $S(\mathbf{q=M},\omega)$. This indicates that the Knight shift on the Oxygen site is not a sensitive probe of such a prominent spectral feature of the spin-$\frac{1}{2}$ KAFH. We suggest to perform Knight shift measurement directly on the Cu site to resolve this issue.

\section{Conclusions and Outlooks}
 
In this work, we have performed a systematic variational Monte Carlo study on the ground state and spin fluctuation spectrum of the spin-$\frac{1}{2}$ KAFH in the RVB theory framework. We find that the best RVB state for the spin-$\frac{1}{2}$ KAFH is described by a $Z_{2}$ gapped mean field ansatz, which hosts a mean field spinon dispersion very different from that of the $U(1)$ Dirac spin liquid state(denoted in this work as $U(1)$-NN state) originally studied in Ref.[\onlinecite{Ran}]. However, we find that the two states are actually very close to each other in the Hilbert space as a result of the non-injective nature of the mapping between the mean field ansatz and the RVB state around the $U(1)$-NN state. We find that such a singular behavior is deeply related to the unique flat band physics on the Kagome lattice and signals the failure of the RVB mean field theory for the spin-$\frac{1}{2}$ KAFH.  
 
Going beyond the RVB mean field theory, we show with the GRPA theory that the spin fluctuation spectrum above the $Z_{2}$ RVB state is almost identical to that above the $U(1)$-NN state and is actually gapless. We find that the spin fluctuation spectrum of the spin-$\frac{1}{2}$ KAFH is not at all featureless, but is characterized by a prominent spectral peak around the $\mathbf{M}$ point at low energy, which contains about half of the total spin fluctuation spectral weight in that momentum region. Interestingly, we find that such an intense spectral peak, which has an energy $\omega_{0}\approx0.25J$ in the thermodynamic limit, lies within a gapless continuum extending to $2.7J$. Such a spectral characteristic is found to agree well with the prediction of recent dynamical DMRG simulation on the spin-$\frac{1}{2}$ KAFH. We thus believe that the intense spectral peak at $\omega_{0}$ should be a robust feature in the spin fluctuation spectrum of the spin-$\frac{1}{2}$ KAFH. 

We argue that the spectral peak below 2 meV in the INS spectrum of Hebertsmithite ZnCu$_{3}$(OH)$_{6}$Cl$_{2}$ should be attributed to the intrinsic spin fluctuation of the Kagome layer, or, more specifically, the prominent spectral peak we found around the $\mathbf{M}$ point, rather than the fluctuation of Cu$^{2+}$ impurity spins occupying the Zn$^{2+}$ site between the Kagome layers. A smaller(or even zero) value of the observed peak energy $\omega_{0}$ as compared to the theoretical prediction(according to which $\omega_{0}\approx0.25J$) implies that Hebertsmithite ZnCu$_{3}$(OH)$_{6}$Cl$_{2}$ is much closer to the $\mathbf{q}=0$ magnetic ordering instability than we thought before. We show that the Knight shift on the Oxygen site is blind to such an intense spectral peak around the $\mathbf{M}$ point as a result of strong antiferromagnetic correlation between nearest neighboring spins on the Kagome lattice. We propose to use the Knight shift on the Cu site as a direct probe of this important spectral feature of the spin-$\frac{1}{2}$ KAFH.

The results presented in this work constitute a concrete example in which the RVB mean field theory fails even at a qualitative level. At the same time, it demonstrates once more the power of the GRPA theory in describing the dynamical properties of the quantum magnet systems. However, giving the fact that the $Z_{2}$ and the $U(1)$-NN state exhibit almost identical excitation behavior in the spin triplet channel, one can not help asking if there exists any qualitative difference in the excitation behavior of the two states in the spin singlet channel. In particular, does the $Z_{2}$ RVB state host a gapped gauge fluctuation spectrum, or, as in the case of the $U(1)$-NN state, is gapless in the gauge channel? We leave such a question to future study.

We acknowledge the support from the National Natural Science Foundation of China(Grant No. 11674391), the Research Funds of Renmin University of China(Grant
No.15XNLQ03), and the National Program on Key Research Project(Grant No.2016YFA0300504). We also thank Ji-Quan Pei for his contribution in the early stage of this work.


\begin{thebibliography}{99}

\bibitem{Elser}V. Elser, Phys. Rev. Lett. \textbf{62}, 2405 (1989).
\bibitem{Chalker}J. T. Chalker and J. F. Eastmond, Phys. Rev. B \textbf{46}, 14201 (1992).
\bibitem{Leung}P. W. Leung and V. Elser, Phys. Rev. B \textbf{47}, 5459 (1993).
\bibitem{Young}N. Elstner and A. P. Young, Phys. Rev. B\textbf{ 50}, 6871 (1994).
\bibitem{Lecheminant}P. Lecheminant, B. Bernu, C. Lhuillier, L. Pierre, and P. Sindzingre, Phys.Rev.B \textbf{56}, 2521 (1997).
\bibitem{Sindzingre}P. Sindzingre and C. Lhuillier, Europhys. Lett. \textbf{88}, 27009 (2009).
\bibitem{Nakano}H. Nakano and T. Sakai, J. Phys. Soc. Jpn. \textbf{80}, 053704 (2011).
\bibitem{Lauchli}A. M. L\"auchli, J. Sudan, and E. S. S\o rensen, Phys. Rev. B \textbf{83}, 212401 (2011).
\bibitem{Series}R. R. P. Singh and D. A. Huse, Phys. Rev. Lett. \textbf{68}, 1766 (1992); 
R. R. P. Singh and D. A. Huse, Phys. Rev. B \textbf{76},
180407(R) (2007); R. R. P. Singh and D. A. Huse, Phys. Rev. B \textbf{77},144415 (2008).
\bibitem{Vidal} G. Evenbly and G. Vidal, Phys. Rev. Lett. \textbf{104}, 187203 (2010).
\bibitem{Singlet} C. Waldtmann, H. U. Everts, B. Bernu, C. Lhuillier, P. Sindzingre, P. Lecheminant, and L. Pierre, Eur. Phys. J. B \textbf{2}, 501 (1998); P. Sindzingre, G. Misguich, C. Lhuillier, B. Bernu, L. Pierre, Ch. Waldtmann, and H. U. Everts, Phys. Rev. Lett. \textbf{84}, 2953 (2000); G. Misguich and B. Bernu, Phys. Rev. B, \textbf{71}, 014417(2005); A. M. L\"{a}uchli and C. Lhuillier, arXiv:0901.1065.
\bibitem{Mila}F. Mila, Phys. Rev. Lett. \textbf{81},2356(1998); M. Mambrini and F. Mila, Eur. Phys. J. B \textbf{17}, 651 (2000).
\bibitem{Auerbach}R. Budnik and A. Auerbach, Phys. Rev. Lett.\textbf{ 93}, 187205 (2004).
\bibitem{Poilblanc}D. Poilblanc, M. Mambrini, and D. Schwandt, Phys. Rev. B \textbf{81}, 180402 (2010).
\bibitem{Sheng}S. S. Gong, W. Zhu, and D. N. Sheng, Sci. Rep. \textbf{4}, 6317 (2014).
\bibitem{He1}Y. C. He, D. N. Sheng and Y. Chen, Phys. Rev. Lett. \textbf{112}, 137202 (2014). 
\bibitem{Changlani}H. J. Changlani, D. Kochkov, K. Kumar, B. K. Clark and E. Fradkin, Phys. Rev. Lett. \textbf{120}, 117202 (2018).

\bibitem{Mendels}P. Mendels, F. Bert, M. A. de Vries, A. Olariu, A. Harrison, F. Duc, J. C. Trombe, J. S. Lord, A. Amato, and C. Baines, Phys. Rev. Lett. \textbf{98} 077204(2007).
\bibitem{Helton}J. S. Helton, K. Matan, M. P. Shores, E. A. Nytko, B. M. Bartlett, Y. Yoshida, Y. Takano, A. Suslov, Y. Qiu, J. H. Chung, D. G. Nocera, and Y. S. Lee,  Phys. Rev. Lett. \textbf{98} 107204(2007).
\bibitem{Han}T. H. Han, J. S. Helton, S. Chu, D. G. Nocera, J. A. Rodriguez Rivera, C. Broholm, and Y. S. Lee, Nature \textbf{492} 406(2012).
\bibitem{Fu}M. X. Fu, T. Imai, T. H. Han and Y. S. Lee, Science \textbf{350} 655(2015).
\bibitem{Shi}Z. L. Feng, Z. Li, X. Meng, W. Yi, Y. Wei, J. Zhang, Y. C. Wang, W. Jiang, Z. Liu, S. Y. Li, F. Liu, J. L. Luo, S. L. Li, G. Q. Zheng, Z. Y. Meng, J. W. Mei, Y. G. Shi
Chin. Phys. Lett.  \textbf{34} 077502(2017).
\bibitem{Mendels}P. Khuntia, M. Velazquez, Q. Barthélemy, F. Bert, E. Kermarrec, A. Legros, B. Bernu, L. Messio, A. Zorko and P. Mendels, Nat. Phys. \textbf{16} 469(2020).


\bibitem{Hastings}M. B. Hastings, Phys. Rev. B \textbf{63}, 014413 (2000).
\bibitem{Ran}Y. Ran, M. Hermele, P. A. Lee, and X. G. Wen, Phys. Rev. Lett. \textbf{98}, 117205 (2007).
\bibitem{Iqbal1} Y. Iqbal, F. Becca, and D. Poilblanc, Phys. Rev. B \textbf{83}, 100404(R) (2011).
\bibitem{Iqbal2} Y. Iqbal, F. Becca, and D. Poilblanc, Phys. Rev. B \textbf{84}, 020407(R) (2011); New J. Phys. \textbf{14}, 115031 (2012); Y. Iqbal, F. Becca, S. Sorella, and D. Poilblanc, Phys. Rev. B \textbf{87}, 060405(R) (2013); Y. Iqbal, D. Poilblanc, and F. Becca, Phys. Rev. B \textbf{89}, 020407(R) (2014).
\bibitem{Iqbal3} Y. Iqbal, D. Poilblanc, and F. Becca, Phys. Rev. B \textbf{91}, 020402(R) (2015).
\bibitem{Liao}H. J. Liao, Z. Y. Xie, J. Chen, Z. Y. Liu, H. D. Xie, R. Z. Huang, B. Normand, and T. Xiang, Phys. Rev. Lett. \textbf{118}, 137202 (2017). 
\bibitem{He}Y. C. He, M. P. Zaletel, M. Oshikawa and F. Pollmann, Phys. Rev. X \textbf{7}, 031020 (2017).
\bibitem{Jiang}S. H. Jiang, P. Kim, J. H. Han and Y. Ran, SciPost Phys. \textbf{7}, 006 (2019).
\bibitem{Zhu}W. Zhu, S. S. Gong, and D. N. Sheng, PNAS, \textbf{116}, 5437 (2019).
\bibitem{Jiang1}H. C. Jiang, Z. Y. Weng, and D. N. Sheng, Phys. Rev. Lett. \textbf{101}, 117203 (2008).
\bibitem{Yan}S. Yan, D. A. Huse, and S. R. White, Science \textbf{332}, 1173 (2011).
\bibitem{Depenbrock}S. Depenbrock, I. P. McCulloch, and U. Schollw\"ock, Phys. Rev. Lett. \textbf{109}, 067201 (2012).
\bibitem{Jiang2}H. C. Jiang, Z. Wang, and L. Balents, Nat. Phys. \textbf{8}, 902 (2012).  
\bibitem{Kolley}F. Kolley, S. Depenbrock, I. P. McCulloch, U. Schollw\"ock, V. Alba, Phys. Rev. B \textbf{91}, 104418 (2015).
\bibitem{Gong}S. S. Gong, W. Zhu, L. Balents, and D. N. Sheng, Phys. Rev. B \textbf{91}, 075112 (2015).
\bibitem{Wen}J. W. Mei, J. Y. Chen, H. He, X. G. Wen, Phys. Rev. B \textbf{95}, 235107 (2017).
\bibitem{Lu}Y. M. Lu, Y. Ran, and P. A. Lee, Phys.Rev.B \textbf{83}, 224413 (2011).
\bibitem{Wen1}X. G. Wen, Phys. Rev. B \textbf{65} 165113 (2002).
\bibitem{Tao1}Tao Li, arXiv:1601.02165.
\bibitem{Iqbal4}Y. Iqbal, D. Poilblanc and F. Becca, arXiv:1606.02255. 
\bibitem{Iqbal5}Y. Iqbal, D. Poilblanc, R. Thomale and F. Becca,  Phys. Rev. B \textbf{97}, 115127 (2018).
 \bibitem{Tao2}Tao Li, arXiv:1805.07689.
 
 
 \bibitem{Li}T. Li and F. Yang, Phys. Rev. B \textbf{81}, 214509 (2010).
 \bibitem{Li1}F. Yang and T. Li, Phys. Rev. B \textbf{83}, 064524(2011).
 \bibitem{Piazza} B. Dalla Piazza, M. Mourigal, N. B. Christensen, G. J. Nilson, P. Tregenna-Piggott, T. G. Perring, M. Enderle, D. F. McMorrow, D. A. Ivanov and H. M. R${\o}$nnow, \emph{Nat. Phys.} \textbf{11}, 62(2015).
 \bibitem{Mei1}J. W. Mei and X. G. Wen, arXiv:1507.03007.
\bibitem{Ferrari}F. Ferrari, A. Parola, S. Sorella, and F. Becca, Phys. Rev. B \textbf{97}, 235103 (2018).
\bibitem{Ferrari1} F. Ferrari and F. Becca, Phys. Rev. B \textbf{98}, 100405(R) (2018).
 \bibitem{Becca}F. Ferrari and F. Becca, Phys. Rev. X \textbf{9}, 031026 (2019).
\bibitem{Ido}K. Ido, M. Imada, and T. Misawa, Phys. Rev. B \textbf{101}, 075124 (2020).
\bibitem{Becca1}F. Ferrari and F. Becca, Phys. Rev. B \textbf{102}, 014417 (2020)


 
\bibitem{Sorella}We are grateful to S. Sorella for pointing out to us this fact.
 \bibitem{note}We note that the spin fluctuation spectrum of the $U(1)$-NN state has been calculated earlier with the GRPA theory in Ref.[\onlinecite{Mei1}]. However, the spectral weight distribution they found is very different from what we presented here. We think that such a difference may originates from the much larger broadening they have used($\delta=0.15J$ as compared to $\delta=0.01J$ used in our calculation), which may have masked the important spectral signature of the spin-$\frac{1}{2}$ KAFH. We note that the cluster size we have used(the largest cluster we have attempted has $24\times24\times3$ sites) is also much larger than that used in Ref.[\onlinecite{Mei1}].    
\end{thebibliography}
\end{document}